\documentclass[acmsmall]{acmart}
\AtBeginDocument{%
  \providecommand\BibTeX{{%
    \normalfont B\kern-0.5em{\scshape i\kern-0.25em b}\kern-0.8em\TeX}}}

\setcopyright{acmlicensed}
\copyrightyear{2018}
\acmYear{2024}
\acmDOI{XXXXXXX.XXXXXXX}

\acmJournal{PACMMOD}
\acmVolume{xx}
\acmNumber{xx}
\acmArticle{xx}
\acmMonth{8}

\newcommand{\rev}[1]{{#1}}

\newcommand{\stitle}[1]{\vspace{1.5ex}\noindent{\bf #1}}
\usepackage{subfigure}
\usepackage{array}
\usepackage{multirow}
\usepackage{epstopdf}
\usepackage[ruled,vlined]{algorithm2e}
\usepackage{url}
\usepackage{makecell}
\usepackage{setspace}
\usepackage{algpseudocode}
\usepackage{url}

\begin{document}

\title{GTS: GPU-based Tree Index for Fast Similarity Search}



\author{Yifan Zhu}
\affiliation{%
  \institution{Zhejiang University}
  \country{China}
}
\email{xtf\_z@zju.edu.cn}

\author{Ruiyao Ma}
\affiliation{%
  \institution{Zhejiang University}
  \country{China}
}
\email{ryma@zju.edu.cn}

\author{Baihua Zheng}
\affiliation{%
  \institution{Singapore Management University}
  \country{Singapore}
}
\email{bhzheng@smu.edu.sg}

\author{Xiangyu Ke}
\affiliation{%
  \institution{Zhejiang University}
  \country{China}
}
\email{xiangyu.ke@zju.edu.cn}

\author{Lu Chen}
\affiliation{%
  \institution{Zhejiang University}
  \country{China}
}
\email{luchen@zju.edu.cn}

\author{Yunjun Gao}
 \affiliation{%
  \institution{Zhejiang University}
  \country{China}
}
\email{gaoyj@zju.edu.cn}

\renewcommand{\shortauthors}{Yifan Zhu, et al.}

\begin{abstract}
Similarity search, the task of identifying objects most similar to a given query object under a specific metric, has gathered significant attention due to its practical applications. However, the absence of coordinate information to accelerate similarity search and the high computational cost of measuring object similarity hinder the efficiency of existing CPU-based methods. Additionally, these methods struggle to meet the demand for \rev{high throughput data management}. To address these challenges, we propose {\sf GTS}, a GPU-based tree index designed for the parallel processing of similarity search in general metric spaces, where only the distance metric for measuring object similarity is known. The {\sf GTS} index utilizes a pivot-based tree structure to efficiently prune objects and employs list tables to facilitate GPU computing. To efficiently manage concurrent similarity queries with limited GPU memory, we have developed a two-stage search method that combines batch processing and sequential strategies to optimize memory usage. The paper also introduces an effective update strategy for the proposed GPU-based index, encompassing streaming data updates and batch data updates. Additionally, we present a cost model to evaluate search performance. Extensive experiments on \rev{five} real-life datasets demonstrate that {\sf GTS} achieves efficiency gains of up to two orders of magnitude over existing CPU baselines and up to 20$\times$ efficiency improvements compared to state-of-the-art GPU-based methods.

\end{abstract}



\received{October 2023}
\received[revised]{January 2024}
\received[accepted]{February 2024}

\maketitle

\section{Introduction}
\label{sec:intro}

Similarity search constitutes a fundamental challenge in the realms of information retrieval and data mining, involving the efficient identification of objects in a dataset that are most similar to a given query object~\cite{ChavezPromximity}. This process carries profound significance across diverse domains, including {multi-media retrieval, decision making, and visualization recommendation}~\cite{pacmmod/LuoZ00CS23, tkde/WangWZLJXCSCZ22,chi/CaiRHHKSWVCST19,pvldb/LeeTABCKMSYHP21}. Operating within metric spaces, similarity search quantifies object similarity via distance metrics, enabling efficient high-dimensional data management across a wide spectrum of data types~\cite{chen2020indexing}. 

\rev{The recent research~\cite{jiang2022big} highlights the pressing need for a general-purpose database capable of handling diverse cancer omics data, spanning molecular omics, molecular interaction, imaging, textual data, etc. While existing solutions like {\sf R-Trees}~\cite{sigmod/Guttman84} excel in handling spatial data, and graph-based methods~\cite{pvldb/WangXY021} are tailored for high-dimensional vector data, the metric space offers a flexible framework for managing {\em dynamic} data of {\em various types} with {\em distinct measures}, such as word cosine distance for semantic measurement and edit distance for DNA strings~\cite{tomek2021promiscuous}. Similarity search in metric spaces is designed with this versatility in mind, providing a general-purpose solution. In addition, metric indexes serve as pivotal solutions for enhancing the scalability of vector databases and offering cost-effective updates for social media and online commercial databases, addressing key issues identified in recent advancements~\cite{is/ShimomuraOVK21,pvldb/WangXY021}, such as the inevitably heavy construction cost and update complexities for graph-based methods.}


Despite its importance, similarity search faces several challenges in achieving efficiency. The absence of coordinate information in metric spaces restricts the utilization of geometric structure knowledge of data points~\cite{jda/MaoMM12}. Additionally, the computation of distances between objects, like edit distance~\cite{pacmmod/ZengTC23}, incurs significant computational overhead. Existing CPU-based methods, such as {\sf MVP-tree}~\cite{sigmod/BozkayaO97,tods/BozkayaO99}, {\sf Spacing-filling curve and Pivot-based B$^+$-tree}~\cite{tkde/ChenGLJC17}, and {\sf EGNAT}~\cite{iccS/MarinUB07,is/NavarroP11}, often fall short in meeting the demands of efficient similarity search. 
\rev{Furthermore, recent technological advancements have heightened the demand for managing and querying large-scale dynamic data, as seen in the exponential growth of DNA data generation facilitated by sequencing technologies~\cite{sigmod/GuptaYCKEMTS21}. These evolving requirements render CPU-based methods insufficient to meet the timely query needs of similarity search.}

Recent advancements in similarity search have turned towards GPU-accelerated methods~\cite{icde/ZhouGJKLLTYZ18}. This shift 
is primarily attributed to the potential parallelism of independent and simple calculations, which has shown promising outcomes~\cite{tjs/BarrientosRHNS22,tkde/LewisT22}.
These methods can be categorized into {\em distance table-based} and {\em tree index-based} methods.
Distance table-based methods~\cite{tjs/BarrientosRHNS22,kim2022es4d,ics/DonnellyG20,icde/Li0QHDY18,tbd/JohnsonDJ21,sc/GaihreZWLSDLL21} employ table structures to facilitate concurrent calculations across all objects, followed by subsequent filtering of unnecessary objects. While GPUs exhibit remarkable concurrent performance enhancements in similarity search compared to CPU-based methods, 
this approach necessitates a significant number of {\em unnecessary distance computations} between {\em all} objects and the query.

On the other hand, tree index-based methods~\cite{tc/KimLC18,tkde/LewisT22,tsc/KimLC22} construct tree-structured indexes to enhance pruning efficiency and utilize GPU cores for concurrent searches {\em on various trees}. However, GPUs face certain limitations in terms of individual computational capacity and core diversity compared to their CPU counterparts~\cite{sigmod/ShahvaraniJ16}.
As a result, current GPU-based methods struggle to achieve {\em concurrent traversal of each tree index}, thus failing to fully exploit the computation potential of GPUs. Consequently, their efficiency improvements over CPU-based methods remain limited.

In this context, some researchers have explored GPU-based approximate similarity search~\cite{icde/ZhouGJKLLTYZ18,icde/YuWZQZL22,icde/ZhaoTL20}. However, they fail to support exact query-answering and lack versatility when it comes to handling a wide range of distance metrics. Our goal is to develop an effective GPU-based index {\em optimized to harness the full computational potential of GPUs while avoiding unnecessary distance computations}. We aim to achieve efficient \rev{batch} similarity search {\em in general metric spaces}, where the only available information is the distance metric function. Nevertheless, to achieve this goal, we must confront and overcome three key challenges.


\vspace{2mm}
{\bf\em Challenge I: Performing data pruning in parallel on a tree index poses significant difficulties in GPU-based parallelization.} CPU-based approaches efficiently implement tree-based indexes through by-level pruning methods to enhance query efficiency by reducing unnecessary distance computations. However, parallelizing this process on GPUs is challenging due to inherent architectural limitations, as elaborated below: 
{\bf (1)} {\em Top-Down Traversal}: \rev{Pruning in the tree structure necessitates a top-down traversal from the root to the leaves, where calculations in succeeding levels depend on those performed at the current level. This interdependency poses a significant hurdle to effective parallelization across levels.}
{\bf (2)} {\em Non-Contiguous Memory Allocation}: In tree structures, nodes at the same level are often stored in non-contiguous memory locations. Consequently, during GPU parallel computation, a linear traversal is necessary to allocate different nodes for concurrent processing. Unfortunately, this allocation process cannot be parallelized efficiently.
{\bf (3)} {\em Limited GPU Core Capabilities}: \rev{GPU cores are typically optimized for parallelizing simple and similar tasks. Conversely, the tree structure may consist of multiple data objects that demand distinct operations. This disparity makes it challenging to efficiently parallelize such diverse tasks on GPUs~\cite{tc/KimLC18}.}

To address these challenges, this paper proposes a novel strategy aimed at {\em reorganizing tree structures within tables} to optimize GPU resources during the pruning process. Initially, a tree-based index structure is built using iteratively chosen pivots, and data partitioning methods are employed to balance the tree and control its height, thereby reducing wastage of GPU resources resulting from the top-down search approach. Subsequently, table-based strategies are utilized to store the indexed tree, allowing consecutive storage of tree nodes at the same level. Such organization enables the parallel allocation of simple GPU tasks for {non-continuous} nodes at the same level, effectively optimizing GPU computing resources and avoiding GPU thread waiting issues caused by serial task allocation.

Additionally, we have designed a {\em cost model} to enhance the parallel computation efficiency of nodes at different levels. The performance of similarity search is closely tied to both object pruning capability and parallel computing efficiency, directly impacted by {\em the node capacity}. A higher node capacity implies a lower tree height, limiting pruning capability but enabling more GPU cores 
for parallel computation of child nodes. Thus, by utilizing the proposed cost model to optimize node capacity, we achieve a well-balanced trade-off between pruning capability and parallel computing efficiency, significantly improving search performance. This improvement is also validated by experimental results.

\vspace{2mm}
{\bf \em Challenge II: Enhancing memory management for \rev{batch} similarity queries.}
\rev{Batch} similarity queries often involve independent query objects, \rev{demanding a considerable amount of memory to store intermediate results. However, surpassing the GPU's memory limit can trigger the memory deadlock issue, which obstructs the release of occupied memory for further computation.} To address this, we propose a {\em two-stage hierarchical query strategy}. In the first stage, batches of concurrent queries are processed in upper-level tree nodes within the index. When the storage capacity of intermediate results approaches a predefined limit, we introduce a node-priority concurrent query strategy, granting access to lower-level tree nodes. This strategy involves grouping the intermediate results of each query and subsequently processing each group sequentially to obtain real results, with the queries in each group computed in parallel. By adopting this approach, we effectively circumvent memory deadlocks and enable high-concurrency similarity queries within our proposed search method.

\vspace{2mm}
{\bf \em Challenge III: Supporting efficient data updates.}
Existing CPU-based approaches rely on pre-computed distances between data objects and pivot or cluster centers for data pruning~\cite{chen2020indexing}. However, dynamic updates to the dataset can impact the index structure, potentially reducing query performance. For instance, a large number of object insertions may alter the overall object distribution, rendering the existing index unsuitable for capturing the characteristics of the updated dataset. Moreover, GPU cores face constraints in supporting dynamic data space allocation~\cite{ppopp/WinterPMS21}, posing challenges in allocating space for newly created tree nodes and modifying existing tree structures in GPU-based tree indexes.
To address these challenges, we take into account the unique characteristics of GPUs, which offer powerful computing capabilities but lack dynamic space allocation abilities. Our proposed solution leverages a {\em compact, high-speed cache table} to implement peak-valley update operations. For low-frequency object updates, we directly update the data in this cache table, avoiding the need to modify the storage structure of the GPU-based tree index. When the cache table exceeds a given size constraint, we perform {\em batch updates} to transfer the data into the tree index. In scenarios involving extensive batch data updates, we adopt an {\em index reconstruction strategy}, capitalizing on the GPU's superior parallel computing performance, which offers a time complexity of $O(\log^3 n)$. As a result, for substantial data updates, we opt to rebuild the tree index entirely. This approach not only enables efficient batch data updates but also guarantees that data updates do not compromise the query efficiency.

In summary, our key contributions are as follows:
\begin{itemize}
\setlength{\itemsep}{-\itemsep}
\item{}  \textit{GPU-based indexing.} We introduce {\sf GTS}, a \underline{G}PU-based \underline{T}ree Index for \underline{S}imilarity Search, which efficiently handles metric range queries and metric \textit{k}-nearest neighbor queries.

\vspace{1mm}
\item{} \textit{Tree-structured index with table-based storage.} We devise a novel pivot-based tree index with a table-based index structure, enabling parallelized calculations of {non-continuous} tree nodes at the same level {for the first time}.

\vspace{1mm}
\item{} \textit{Concurrent similarity search.} We develop concurrent search methods for our proposed GPU-based-tree index to prevent memory deadlock. Additionally, we introduce a cost model that balances concurrency and pruning capability.

\vspace{1mm}
\item{} \textit{Dynamic updates.} We propose effective update strategies tailored for dynamic scenarios, including both streaming data updates and batch data updates.

\vspace{1mm}
\item{} \textit{Extensive experiments.} We conduct comprehensive experimental evaluations on \rev{five} real datasets, demonstrating the remarkable efficiency gains achieved by {\sf GTS}. It outperforms existing CPU baselines by up to two orders of magnitude and even surpasses GPU-based general methods by up to 20$\times$ in terms of efficiency. 

\end{itemize}

The rest of this paper is organized as follows. We review the previous works in Section~\ref{sec:relatedwork}, and present the problem statement in Section~\ref{sec:problemformulation}. The GPU-based tree index is introduced in Section~\ref{sec:Index}, and the similarity search process with the cost model is detailed in Section~\ref{sec:Search}. Comprehensive experimental studies and our findings are reported in Section~\ref{sec:exp}. Finally, Section~\ref{sec:conclusion} concludes the paper and offers directions for future research.

\section{Related Work}
\label{sec:relatedwork}

\stitle{CPU-based Similarity Search.}
CPU-based similarity search in metric spaces has received significant attention. Existing approaches can be categorized into {\em table-based} methods, {\em tree-based} methods, and {\em hybrid} methods that combine the previous two groups of approaches.
Table-based methods pre-compute the {\em full} set of element-wise distances between pivots (reference points) and objects, manage them in a large table for quick access, and utilize triangle inequality to narrow down search spaces. Prominent examples include {\sf List of Clusters}~\cite{spire/ChavezN00,prl/ChavezN05}, {\sf LAESA}~\cite{prl/MicoOV94}, and {\sf EPT}~\cite{sisap/RuizSCFT13}. On the other hand, tree-based methods, including {\sf BST}~\cite{tse/KalantariM83}, {\sf M-tree}~\cite{vldb/CiacciaPZ97}, {\sf MVP-tree}~\cite{sigmod/BozkayaO97,tods/BozkayaO99}, {\sf spacing-filling curve, and pivot-based B$^+$-tree}~\cite{tkde/ChenGLJC17}, employ hierarchical tree structures to manage objects in a top-down manner, which rapidly identifies candidate items that meet the search criteria, thereby reducing the search space and improving query efficiency.
Hybrid methods aim to leverage the advantages of both table-based and tree-based approaches. For instance, {\sf CPT}~\cite{sisap/MoskoLS11} links the distance table with M-tree leaves, while {\sf GNAT}~\cite{vldb/Brin95} and {\sf EGNAT}~\cite{iccS/MarinUB07,is/NavarroP11} store the distance table of the minimum bounding box in tree nodes. These methods combine pre-computed distance information with hierarchical data structures, resulting in high single-point query performance, making them suitable for a wider range of similarity search tasks.

Despite the progress made in CPU-based similarity search methods, they face limitations~\cite{sigmod/ShahvaraniJ16}. A primary challenge is the {\em limited parallelism} offered by CPUs, which are predominantly designed for sequential processing. Furthermore, the {\em computational intensity} of search operations can strain CPU performance, making real-time query processing for large-scale datasets a challenging task. To address the increasing demand for higher search performance, recent studies have turned to GPU-accelerated approaches. Capitalizing on the inherent advantages of GPUs, such as massive parallelism and superior computational power, GPU-based methods have the potential to overcome the shortcomings of CPU-based techniques. They unlock more efficient concurrent similarity search and significantly improve query throughput.

\stitle{GPU-accelerated Table-based Methods.}
By utilizing table structures, we can efficiently perform similarity calculations between database objects and the query concurrently, enabling the parallel filtering of objects that meet the query conditions. Notable examples like {\sf Pivot-based Heaps}~\cite{tjs/BarrientosRHNS22}, {\sf ES4D}~\cite{kim2022es4d}, and {\sf COSS}~\cite{ics/DonnellyG20} employ pivots, lookup arrays, and bin arrays to efficiently compute distances and enhance search termination or support general similarity search. Additionally, {\sf G-Grid}~\cite{icde/Li0QHDY18} leverages GPU-searchable grid tables for exact \textit{k}-NN queries in road networks, while {\sf GPU-based Batch search}~\cite{tbd/JohnsonDJ21} optimizes brute-force, approximate, and compressed-domain search to facilitate billion-scale similarity search. The recent {\sf Dr. Top-\textit{k} algorithm}~\cite{sc/GaihreZWLSDLL21} introduces delegate-centric concepts and system optimizations to achieve fast multi-GPU top-\textit{k} computation. However, existing table-based methods still suffer from the need to calculate similarity distances for {\em all} database objects and the query, resulting in numerous unnecessary computations that hinder search efficiency.

\stitle{GPU-accelerated Tree-based Methods.}
To address the issue of unnecessary distance computations, researchers have proposed GPU-based tree indexes. For instance, {\sf G-tree}~\cite{tc/KimLC18} combines the efficiency of R-tree in low-dimensional space with a best-first strategy for concurrent tree-based similarity search on GPUs. {\sf G-PICS}~\cite{tkde/LewisT22} aims at parallelized processing by constructing various spatial trees with different partitioning methods. {\sf LBPG-tree}~\cite{tsc/KimLC22} optimizes the usage of multiple GPUs by compacting and sorting candidate tree nodes and optimizing search schedules for concurrent queries. \rev{Aiming at optimizing computing resource utilization, some methods~\cite{gis/YouZG13,gis/TengNEBK021,aspdac/LuoWL12} leverage a level-based concurrent construction approach to build the index.} While these approaches represent significant advancements, challenges persist in concurrent similarity search. \rev{Existing GPU-based tree indexes often employ similarity search by (1) applying fixed-size thread blocks and (2) sequentially processing each tree node. This approach may lead to certain blocks having an excessive number of answers to verify, causing size overflow and potential memory deadlock issues. Conversely, other blocks may remain idle due to the absence of leaf nodes to explore, leading to a significant workload imbalance.}

\vspace{2mm}
In contrast to common CPU-based practices that link tree elements with pre-computed distance table cells, our novel GPU-based tree index, {\sf GTS}, is efficiently {\em maintained within a table}. \rev{This design allows us to concurrently compute non-continuous tree nodes at the same level via sort and coding strategies, and ensures uniform and straightforward computing tasks across all CPU cores, achieving higher efficiency.} Additionally, we introduce two-stage search methods that enhance {\em memory utilization} and a cost model for evaluating {\em node capacity}, resulting in substantial improvements in concurrency and throughput. Furthermore, we have developed effective streaming data updates and batch update strategies to efficiently manage dynamic scenarios. These collective improvements position {\sf GTS} as a powerful and efficient solution for concurrent similarity search tasks on GPU-based architectures.

\section{Problem Formulation}
\label{sec:problemformulation}
We formally introduce the metric space and the similarity search below. 
For clarity, Table~\ref{tab:symbol} summarizes frequently used notations.

\begin{table}\label{tab:symbol}
	\centering
	\renewcommand\arraystretch{1}
	\caption{{Symbols and description}}\vspace{-2mm}
	\label{tab:symbol}
	\small
	\setlength{\tabcolsep}{3pt}
	\begin{tabular}{p{4cm}p{10cm}}
		\hline
		\textbf{Notation} & \textbf{Description} \\
		\hline
        $q$, $o$, $p$ & A query, an object, and a pivot in a metric space\\        
        $O$ & An object set in a metric space\\        
        {$n$} & The number of objects in set $O$ \\        
        $d(\cdot,\cdot)$ & A distance metric\\        
        {$N$} & A tree node \\
        {$N\_list,T\_list$} & The node list and the table list \\        
        $\textit{MRQ}(\cdot, \cdot)$ & A metric range query\\	
        $\textit{MkNNQ}(\cdot, \cdot)$ & A metric $k$ nearest neighbour query\\	
		\hline
	\end{tabular}
	\vspace{-2mm}
\end{table}

A metric space is represented as a tuple $(M, d)$, where $M$ denotes the domain of objects, and $d$ serves as the distance metric used to quantify the similarity between any pair of objects $(o_1, o_2)$ within this space. The distance metric $d$ must adhere to the fundamental principles, including {\em symmetry}, {\em non-negative}, {\em identity}, and {\em triangle inequality}, i.e., $d(o_1, o_2)\leq d(o_1, o_3) + d(o_3, o_2)$. Distance metrics, such as {\sf Manhattan distance}, {\sf Euclidean distance}, and {\sf word edit distance}~\cite{pacmmod/ZengTC23}, satisfy the aforementioned conditions and are commonly employed in metric spaces. An essential advantage of the metric space is its ability to accommodate various data models since it does not impose any requirements on the data type~\cite{chen2020indexing}. 

Building upon these principles, we can formally define two types of metric similarity searches:

\begin{definition}
\label{defn:MRQ}
{\bf (Metric Range Query.)} Given an object set $O$, a query object $q$, and a search radius $r$ in a metric space, a metric range query (MRQ) finds all objects in $O$ that are within a distance $r$ from $q$. Formally, $MRQ(q, r) = \left\{o\vert \ o \in O \land d(q, o) \leq r\right\}$.
\end{definition}

\begin{definition}
\label{defn:kNNQ}
{\bf (Metric $k$ Nearest Neighbor Query.)} Given an object set $O$, a query object $q$, and an integer $k$ in a metric space, a metric $k$ nearest neighbor query (M\textit{k}NNQ) finds a set $S$ of $k$ objects in $O$ that are most similar to $q$. Formally, $MkNNQ(q, k)= \{S\vert \ S \subseteq O \land \vert S\vert = k \land \forall s \in S, o \in O-S$,
$d(q, s) \leq d(q, o)\}$.
\end{definition}

\begin{figure}
  \includegraphics[width=0.77\linewidth]{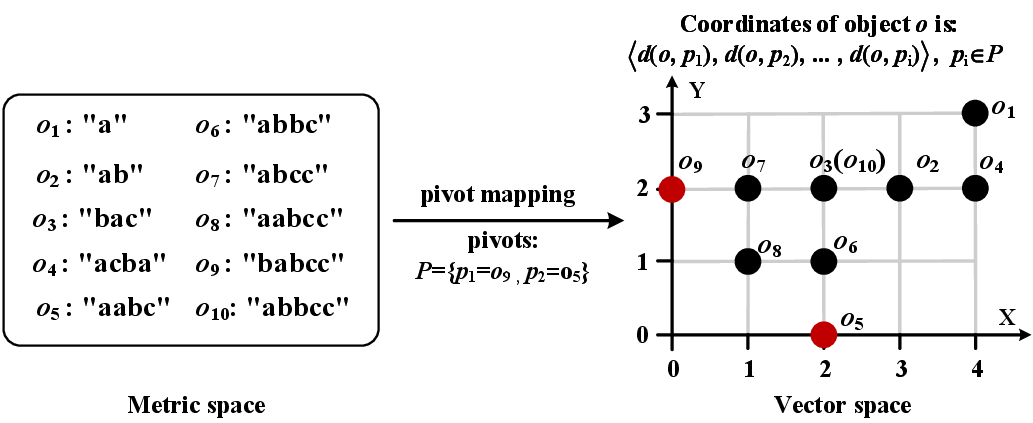}\vspace{-3mm}
  \caption{{Metric space and pivot mapping}}
  \label{fig:m-examp}
\end{figure}

Consider the metric space for a string dataset shown in Fig.~\ref{fig:m-examp},  with object set $O = \{o_1,o_2,...,o_9\}$ and word edit distance $d(\cdot,\cdot)$ measuring similarity, a metric range query $MRQ(o_1,2)$ aims to find objects that can be transformed to $o_1$ within 2 edit operations (insertion, deletion, or replacement). Consequently, $MRQ(o_1,2)$ returns the set $\{o_1, o_2, o_3\}$. A metric 3 ($k=3$) nearest neighbor query $MkNNQ(o_1, 3)$ finds 3 objects from $O$ that can be modified with the least operations to become most similar to the query object $o_1$. Thus, $MkNNQ(o_1, 3) = \{o_1, o_2, o_3\}$. Note that if the distance from the query to its $k$-th nearest neighbor is known in advance, an M\textit{k}NNQ can be efficiently answered by performing an MRQ. For instance, when finding $MkNNQ(o_1, 3)$, we can obtain the answer by performing $MRQ(o_1, 2)$, assuming that the search distance $r=2$ from $o_1$ to its $3^{rd}$ nearest neighbor has been provided in advance.

Indexing a metric space that lacks coordinate information can be challenging, as traditional mathematical tools used in vector spaces cannot be directly applied. To tackle this, a common approach is to select a number of reference points, known as {\em pivots}, to map the data objects into a low-dimensional space~\cite{PSAMSZhu,jda/MaoMM12}. Each chosen pivot serves as an individual dimension, representing the distances from the pivot to all the data objects. This transformation allows the metric space to be represented in a structured form, similar to a vector space, enabling further analysis.

When employing a pivot set, such as $P= \{p_1=o_9, p_2=o_5\}$, to map a metric space into a vector space, each object $o$ is represented as a coordinate $\langle d(o,p_1),d(o,p_2)\rangle$, as depicted in Fig.~\ref{fig:m-examp}. Thus, the distance between the objects $o_1$ and $o_2$ can be approximated by $|d(o_1-p_i)-d(o_2-p_i)|$, using the pivot $p_i$ in the coordinate (dimension) $D_i$ of the mapped vector space, which is smaller than the real distance $d(o_1,o_2)$ according to the triangle inequality.
Therefore, when answering a similarity search, an efficient solution is to first search the mapped vector space with indexes, and then verify each found object to return the real answers. For instance, when answering $MRQ(o_1, 1)$, it is more efficient to search the vector space {along each coordinate} to find objects that are within distance 1 to $o_1$ along any dimension to find the candidate objects $\{o_1,o_2,o_4\}$ for $MRQ(o_1, 1)$ {in the metric space.} 
By verifying each candidate, $\{o_1,o_2\}$ are returned as the real answers. This pivot-based indexing technique effectively reduces the search space and speeds up the search process, making it a valuable solution for efficiently handling metric spaces without explicit coordinate information.

\section{Index}
\label{sec:Index}
In this section, we present an overview of our GPU-based tree index designed for similarity search, along with the structure of our proposed index. Subsequently, we delve into the construction and update strategies employed to support dynamic scenarios effectively. To solidify the understanding of our approach, we also provide a theoretical analysis that covers space consumption and time complexity of the proposed method. 

\subsection{Overview}
\label{sec:Overview}
Unlike CPUs that possess a few powerful cores optimized for handling complex tasks efficiently, GPUs offer abundant computing resources, making them ideal for {\em parallel} processing of numerous {\em simple computational tasks} independently. Therefore, linear continuous structures such as tables are well-suited for the parallel nature of GPUs, as each GPU core can process individual values in the table simultaneously. On the other hand, tree indexes, with hierarchical structures for pruning, can effectively narrow down the search space and accelerate the search process. 

\begin{figure}[t]
    \centering    
    \includegraphics[width=0.6\linewidth]{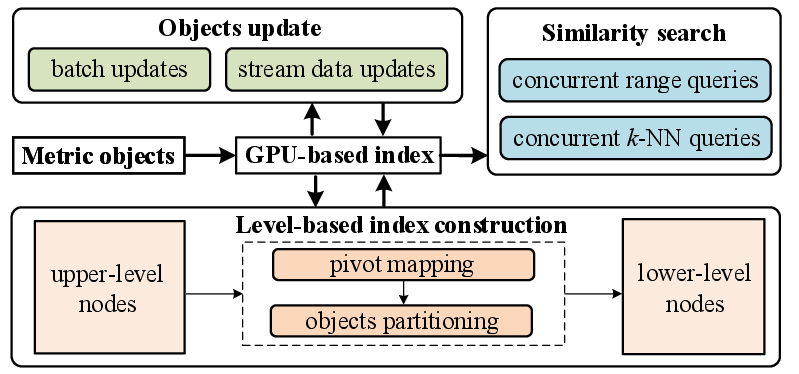}
    \vspace{-2mm}
    \caption{{The overall framework of our work {\sf GTS}}}
    \vspace{-2mm}
    \label{fig:framework}
\end{figure}

Motivated by these factors, we design a GPU-based tree index called {\sf GTS}, which fully {\em harnesses the pruning properties of tree structures} and the {\em parallelism capabilities of GPUs} to achieve efficient similarity search in metric spaces. The framework of {\sf GTS} is depicted in Fig.~\ref{fig:framework}. Firstly, we construct a GPU-based tree index in a hierarchical manner, where upper-level nodes split the objects into lower-level nodes using pivot mapping and object partitioning. To accommodate dynamic scenarios, we design stream data updates and batch updates, catering to different update requirements. Finally, we utilize the GPU-based tree index to concurrently process multiple queries in a batch, significantly accelerating general similarity search in metric spaces, including \rev{range queries and $k$-nearest neighbor ($k$-NN) queries in batches}. 
These techniques effectively exploit the pruning characteristics of the tree index and the parallelism of GPUs to enhance the overall search performance. 

\subsection{Index Structure}
\begin{figure*}[t]
  \vspace{-1mm}
  \includegraphics[width=0.99\linewidth]{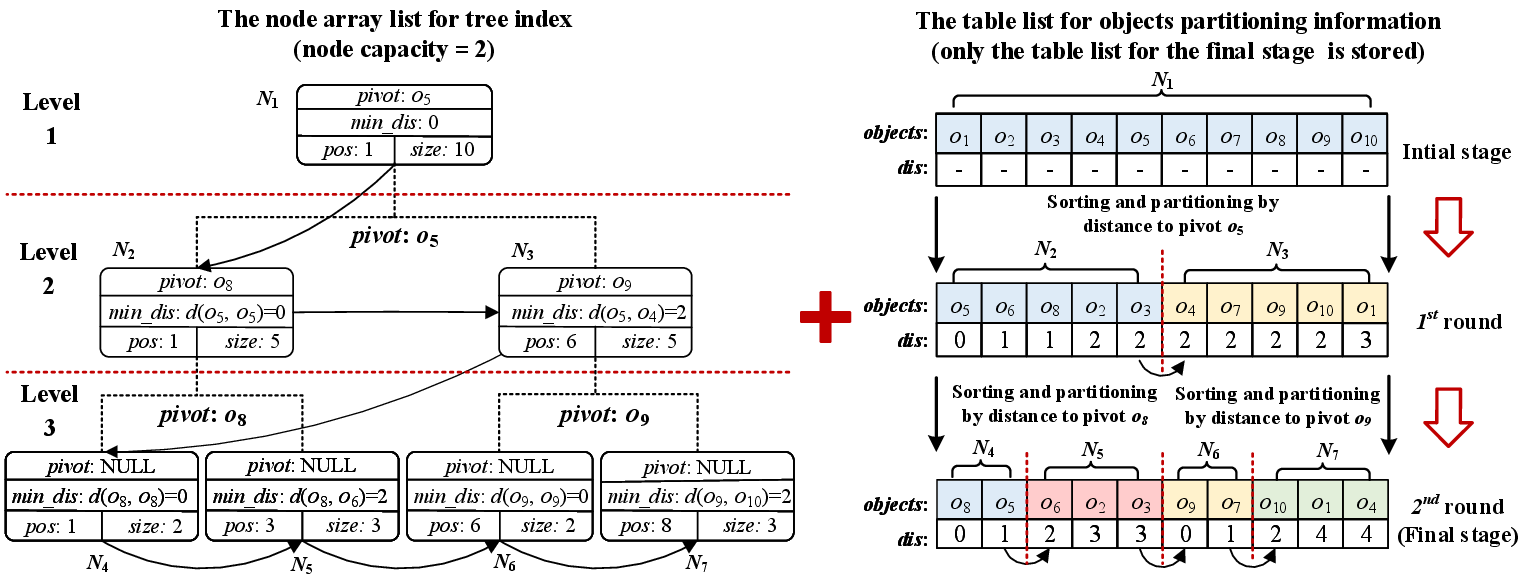}
  
  \vspace{-2mm}
  \caption{\rev{The structure of {our proposed} GPU-based tree index, {\sf GTS}}}
  \vspace{-2mm}
  \label{fig:index_structure}
\end{figure*}

Our proposed GPU-based tree index {\sf GTS} consists of two key components: \textbf{the tree index} and \textbf{the table list}, as illustrated in Fig.~\ref{fig:index_structure}. The balanced tree index efficiently partitions the data objects after pivot mapping, facilitating hierarchical data management and achieving excellent pruning efficiency. To ensure efficient parallel computations on GPUs, we employ a node list structure to manage the tree nodes and store the distances from the pivots to the objects within each node using the table list, which is stored sequentially and contiguously. For a clearer understanding, Fig.~\ref{fig:index_structure} provides an illustrative example of our indexing structure applied to the metric space featured in Fig.~\ref{fig:m-examp}. In the following, we will provide a comprehensive explanation of the detailed structure.

\vspace{2mm}
\noindent
\textbf{The Tree Index.} Similar to the {\sf MVPT}~\cite{sigmod/BozkayaO97,tods/BozkayaO99}, which is renowned as the most efficient CPU-based in-memory metric index~\cite{chen2020indexing}, the tree nodes of our proposed {\sf GTS} also maintain crucial information such as the selected pivot, the minimum distance {from the enclosed objects} 
to the pivot (denoted as $min\_dis$), and the size of the data objects under their management.
To ensure efficient utilization of GPU resources during similarity calculations, we store all the tree nodes in a node table list. Specifically, these nodes are linearly linked, and their ID numbering follows the principles of a full multi-way tree. By designating the ID of the root node as 1, the ID of the $j$-th child node w.r.t. node $N_i$ is determined as follows~\cite{tkde/LewisT22}:
\begin{equation}
ID = (i-1) * N_c + j + 1, 
\end{equation}
where $N_c$ denotes the node capacity, representing the maximum number of child nodes each node has. For example, in Fig.~\ref{fig:index_structure}, each node $N_i$ ($1\leq i <7$) is linked to the next node $N_{i+1}$, and the $j$-th child node of $N_i$ is $N_{(i-1)*N_c+j+1}$, e.g., the second child node of $N_3$ is $N_7$. 
{This design maximizes the GPU resources by enabling the parallelized computation for non-continuous tree nodes at the same level, and will be detailed in Section~\ref{sec:Index_Construction}.} Additionally, to maximize space utilization, we pre-compute the theoretical maximum height \rev{$max\_h=\lceil log_{N_c}{(|O|+1)} \rceil-1$} of the index, where $|O|$ is the total number of the objects. We then set the height bound $h$ to be ($max\_h-1$), making some nodes at the last level overfull and not eligible for splitting. For instance, the size ($=3$) of node $N_7$ exceeds the node capacity ($=2$).

Furthermore, to store the object partitioning information, which includes the objects and their distances to the pivots in each tree node, we maintain the start position of each node in the table list (denoted as $pos$). Let's consider node $N_3$ in Fig.~\ref{fig:index_structure} as an example, where the pivot is $o_9$, the start position of node $N_3$ in the table list is 6, and its $size$ is 5. Thus, according to the second level of the table list, we can deduce that $N_3$ maintains the objects $o_4,o_7,o_9, o_{10}$, and $o_{1}$. It is important to note that these objects are mapped by the pivot $o_5$ in the parent node of $N_3$ and the minimum distance $min\_dis$ from the pivot $o_5$ to all data objects in node $N_3$ is 2.  

\label{sec:Index_Structure}

\vspace{2mm}
\noindent
\textbf{The Table List.} We use a table list to maintain the object partitioning information for each level of the tree. Considering the example shown in Fig.~\ref{fig:index_structure}, at the first level, node $N_1$ contains all ten objects since no data objects are partitioned yet.  Next, using pivot $o_5$ to map objects in $N_1$, $N_1$ is split into $N_2$ and $N_3$. $N_2$ maintains the first five objects with the smallest distances to the pivot, 
while $N_3$ maintains the rest. Therefore, $\{o_5,o_6,o_8,o_2,o_3\}$ are partitioned into $N_2$ in ascending order based on their distances to the pivot $o_5$, while $\{o_4,o_7,o_9,o_{10},o_1\}$ are allocated to $N_3$ in the same manner. Subsequently, the objects in $N_2$ and $N_3$ are partitioned again using pivots $o_8$ and $o_9$ respectively, resulting in the creation of nodes $N_4$ to $N_7$ in the third level. The table list records the partitioning information for all the tree nodes. Notably, since lower-level tree nodes are partitioned from the upper-level nodes, we can obtain the objects' partitioning information of upper-level nodes by combining the table list information of the lower-level objects. For example, the objects partitioning information of node $N_3$ can be derived by combining the list information of its child nodes $N_6$ and $N_7$. Therefore, to optimize memory consumption, we store the table list solely for the last level (i.e., leaf level) of the tree.  

\subsection{Index Construction}
\label{sec:Index_Construction}
{The index construction process in GPU-based solutions involves data transfer from main memory to GPU and the data management via the index maintained within the GPU memory.} Existing GPU-based tree construction strategies often assign each GPU core to construct individual tree {nodes~\cite{jpdc/KimKN13,jocs/Nakasato12}}. However, this approach encounters challenges when dealing with tree nodes that manage a large number of objects.
Since each GPU core can only handle relatively simple computational tasks, splitting such tree nodes becomes problematic. For example, attempting to split the root node into second-layer tree nodes would require storing all the objects in a single GPU core, which is practically infeasible due to {memory limitations}.
To overcome this limitation, some {methods~\cite{iscas/HuNA16,tkde/LewisT22}} build multiple trees, with each tree managing only a small number of objects. While this approach increases parallelism during index construction, it introduces significant space consumption when handling high query throughput. Querying multiple trees for each query object consumes considerable space, thereby restricting the number of queries that can be answered in parallel and causing workload imbalances across the trees.
To address these issues, we propose a top-down construction algorithm based on distance encoding to achieve high-level parallelism. 
{Instead of assigning each GPU core to individual nodes~\cite{jpdc/KimKN13,jocs/Nakasato12}, we construct nodes at the same level simultaneously using all the GPU cores to facilitate parallelism. This is because the distance computation of each node is independent and all nodes at each level are continuously stored. Specifically, we partition the objects following a top-down manner. At each level $i$, we simultaneously map the objects with pivots to all the tree nodes at the current level $i$. Next, we employ a {\em global sort algorithm with distance encoding} to concurrently partition the objects at the level ($i+1$). 
Since all objects are sorted and partitioned together using the existing table stored in global memory, we can fully utilize the GPU cores without being limited by space constraints. Finally, we construct a single tree that maintains all the objects, 
{breaking the construction bottleneck encountered by multi-tree methods~\cite{iscas/HuNA16,tkde/LewisT22}.} 
In the following, we detail the construction of {\sf GTS}. 

\vspace{2mm}
\noindent
\textbf{Index Construction.} Algorithm~\ref{algo:index_construction} outlines the construction process of our index. Firstly, it calculates the tree height (denoted as $h$) based on the dataset size $|O|$ and the node capacity $N_c$ (line 1). Next, the algorithm initializes the root node, the current layer number, the starting position of the root node in the node list, the number of nodes in the current layer, and the table list (lines 2--5). Subsequently, a while loop is executed to construct the index until the current level exceeds the height boundary (lines 6--10). In each iteration, the node list $N\_list$ of the index and the table list $T\_list$ are updated through pivot mapping (line 7) and object partitioning (line 8) operations. Finally, the algorithm completes the construction of {\sf GTS} and returns $N\_list$ and $T\_list$ (line 11).

\begin{algorithm}[t]
\small
\SetNlSty{small}{}{:}
\LinesNumbered
\setstretch{1}
\DontPrintSemicolon
\caption{Index Construction}
\label{algo:index_construction}
\KwIn{an object set $O$, a distance metric $d$, and the node capacity $N_c$ that controls the tree height}
\KwOut{the node list $N\_list$ and the table list $T\_list$}
    $h \gets$ \rev{$\lceil log_{N_c}{(|O|+1)} \rceil-1$}\; 
    $N\_list[1] \gets$ the initialized root node with $size=|O|,pos=0$\; 
    $layer \gets 1$, $pos \gets 1$,  $N_s\gets 1$ // initialize the current layer $layer$, the starting position $pos$ in the node list, and the number $N_s$ of nodes in the current layer\;
    \ForEach{$o_i \in O$ \textbf{in parallel}}
    {
        $T\_list[i].object\gets o_i$ // initialize the table list\;
    }
    \While{$layer\leq h$}{                
        $N\_list,T\_list \gets$\textbf{Mapping}($d,pos,N_s,N\_list,T\_list$ )\;
        $N\_list,T\_list \gets$\textbf{Partitioning}($N_c,pos,N_s,N\_list,T\_list$)\;     
        
        $pos\gets (pos-1)*N_c+2$\;
        $N_s\gets N_c*N_s$, $layer\gets layer+1$\;       
    }
    \KwRet $N\_list,T\_list$\;
\end{algorithm}

\vspace{2mm}
\noindent
\textbf{Pivot Mapping.} We present the GPU-based pivot mapping process in Algorithm~\ref{algo:pivot_mapping}. In this algorithm, we employ the {\sf FFT}~\cite{mor/HochbaumS85} as the pivot selection algorithm, which selects the object farthest from existing pivots as the new pivot and can be easily parallelized by concurrently computing the distances, thereby improving efficiency. \rev{In terms of the initial pivot, several existing pivot selection algorithms, such as {\sf FFT}, {\sf BPS}~\cite{prl/MicoOV94}, and {\sf HF}~\cite{vldb/TrainaFTVF07}, opt for a random object as the first pivot, as they have validated that the initial pivot has minimal impact on search performance. Moreover, the recent study~\cite{PSAMSZhu} concludes that no algorithm exists to select the optimal pivots on any dataset. This makes it impossible to choose an optimal initial pivot with a generalized strategy. Therefore, we also choose the first pivot randomly.}
The pivot mapping algorithm begins by locating the nodes in the current layer from the node list $N\_list$ in parallel (lines 1--2). Next, it retrieves the objects maintained by the current node $N$ from the table list $T\_list$ in parallel (lines 3--4). Then, the pivot is chosen from objects maintained by the temp table $T\_tmp$ using the FFT algorithm, and the distances between the pivot and the objects maintained by $N$ are computed simultaneously (lines 5--7). 
{Additionally, to further improve efficiency, the pivots of each node are stored in the shared memory, which is much faster than GPU global memory~\cite{tkde/LewisT22}.}
Finally, the algorithm returns the updated node list $N\_list$ and table list $T\_list$ (line 8).

\begin{algorithm}[t]
\small
\SetNlSty{small}{}{:}
\LinesNumbered
\setstretch{1}
\DontPrintSemicolon
\caption{Mapping}
\label{algo:pivot_mapping}
\KwIn{a distance metric $d$, the starting position $pos$, the node size $N_s$, the node list $N\_list$, and the table list $T\_list$}
\KwOut{the updated $N\_list$ and $T\_list$}
    \ForEach{$i\in [0, N_s)$ \textbf{in parallel {across blocks}}}{
        $N \gets N\_list[pos+i], T\_tmp\gets\{\varnothing\}$\;         
        \ForEach{$j\in [0, N.size)$ \textbf{in parallel {within block}}}{
                $T\_tmp[j].object \gets$ $T\_list[j+N.pos].object$ \;
        }
        $N.pivot \gets$ the object chosen by FFT~\cite{mor/HochbaumS85} from $T\_{tmp}$\;
        \ForEach{$j \in [0, N.size)$ \textbf{in parallel {within block}}}{            
            $T\_list[j+N.pos].dis\gets  d(T\_{tmp}[j].object,N.pivot)$\;
        }
    }
  \KwRet $N\_list,T\_list$\; 
\end{algorithm}

\vspace{2mm}
\noindent
\textbf{Objects Partitioning.} After completing the pivot mapping process, we obtain the distances between data objects and the pivots in each node at the current level. Subsequently, we partition the objects to the newly constructed tree nodes based on these distances. Managing the objects in each node effectively requires sorting them based on their distances to the pivot. However, performing individual sorts using each GPU core for each node incurs significant time overhead. To address this challenge, we propose a global object partition strategy to efficiently leverage all GPU computing resources. Firstly, we encode the individual distances from each object to the corresponding pivot in parallel. Subsequently, we employ a global concurrent sorting algorithm to arrange all the encoded distances in ascending order. This arrangement groups the objects that should be partitioned to the same newly constructed tree node together. Finally, we decode the distances of each object and construct the tree nodes in the next layer in parallel. \rev{Notably, the objects in each tree node are uniformly divided based on the quantity and are then correctly indexed in the child nodes. Although identical objects with duplicate keys can be partitioned into different nodes, causing the tree nodes to overlap, this strategy ensures a balanced tree structure with efficient and accurate similarity search.}

Algorithm~\ref{algo:objects_partitioning} presents the detailed steps to partition the objects. Firstly, the algorithm normalizes the individual distances derived through pivot mapping in the current layer to the range [0,1) in parallel (lines 1--2). Next, it encodes the distances of each object, considering the information of the node to which the object belongs (lines 3--6). Specifically, given an object $o$ that belongs to the node $N_i$, the distance $dis$ from $o$ to the pivot $N_i.pivot$, and the upper bound $max$ of $dis$, the encoded distance $dis'$ of $o$ is determined as $dis'= \frac{dis}{max+1}+i$.
By doing so, the integer part of $dis'$ records the node belonging information of $o$, while the decimal part of $dis'$ records the distance from $o$ to the corresponding pivot. After applying the sorting algorithm offered by GPU on the encoded distances (line 7), the algorithm partitions the objects in each node concurrently (lines 8--19). {This enables simultaneous computation of non-continuous tree nodes at the same level}. For each node $N$ in the current layer, the algorithm first decodes the distances of each object belonging to $N$ in parallel (lines 10--11). Then, after initializing the newly constructed child nodes $N'$ of $N$ in the next layer, the algorithm distributes the objects evenly into each new node $N'$ (lines 12-18). Finally, the object partitioning completes and the algorithm returns the updated $N\_list$ and $T\_list$ (line 19).

\begin{algorithm}[t]
\small
\SetNlSty{small}{}{:}
\LinesNumbered
\setstretch{1}
\DontPrintSemicolon
\caption{Partitioning}
\label{algo:objects_partitioning}
\KwIn{the node capacity $N_c$, the starting position $pos$, the node size $N_s$, the node list $N\_list$, and the table list $T\_list$}
\KwOut{the updated $N\_list$ and $T\_list$}
    get the maximum distance $max$ in the $T\_list$ \textbf{in parallel}\; 
    normalize each distance $dis$ in the $T\_list$ to $\frac{dis}{max+1}$\;
    \ForEach(\tcp*[h]{encoding}){$i\in [0, N_s)$ \textbf{in parallel {across blocks} }}{
        $N \gets N\_list[pos+i]$\;         
        \ForEach{$j \in [0, N.size)$ \textbf{in parallel {within block} }}{
            $T\_list[j+N.pos].dis \gets$ $i+T\_list[j+N.pos].dis$ \;
        }
    }
    sort all the distances in $T\_list$ using GPU\;
    \ForEach{$i\in [0, N_s)$ \textbf{in parallel {across blocks} }}{
        $N \gets N\_list[pos+i]$\;         
        \ForEach(\tcp*[h]{decoding}){$j\in [0, N.size)$ \textbf{in parallel {within block} }}{
            $T\_list[j+N.pos].dis \gets$ $(T\_list[j+N.pos].dis-i)*(max+1)$ \;
        }
        $avg\_size\gets \lfloor \frac{N.size}{N_c} \rfloor$\;
        \ForEach{$j\in[0, N_c)$ \textbf{in parallel {within block} }}{
             $N' \gets N\_list[(N.pos-1) * N_c + j + 2]$\;  
             $N'.pos \gets N.pos+j*N_c$\;
             \textbf{if}$j<N_c-1$ \textbf{then} $N'.size \gets avg\_size$\;
            \textbf{else} $N'.size \gets N.size-avg\_size*(N_c-1)$\;            
             $N'.min\_dis\gets T\_list[N'.pos].dis$\;
        }
    }
  \KwRet $N\_list,T\_list$\;
\end{algorithm}

\subsection{Index Updating}
\label{sec:Index_Updating}
To cater to dynamic scenarios, we have devised efficient data update strategies for our {\sf GTS} index. Data updates can encompass insertions, deletions, and modifications, where each modification can be regarded as a deletion followed by an insertion. To handle different application scenarios, data updates can take two forms: 1) Real-time or incremental updates are suitable for continuously arriving data streams, such as real-time monitoring and real-time data analysis; and 2) Batch updates are designed to handle large volumes of data updates, typically encountered in scenarios involving database batch management. To effectively accommodate these distinct update scenarios, we introduce two separate index update strategies: {\em Stream Data Updates} and {\em Batch Updates}.

\vspace{2mm}
\noindent
\textbf{Stream Data Updates.} 
When data arrives in a sequential streaming fashion, traditional data updates for tree index can lead to {\em structure changes}, potentially affecting search efficiency and incurring a logarithmic time cost to update corresponding tree nodes. 
{Nonetheless, incrementally updating indexes can be time-consuming on GPU, whose architecture fails to support efficient individual operations~\cite{sigmod/ShahvaraniJ16}.}
\rev{
Motivated by the {\sf LSM-Tree} update strategy that reduces write amplification by buffering writes in memory and subsequently merges them using sequential operations~\cite{vldb/LuoC20}, we propose {\em a lazy strategy with a cache list} to handle streaming data updates. This approach aims to harness the superior processing power of GPUs.}
Specifically, when new objects arrive, instead of directly updating the index, we store the newly inserted data in a cache list. For data deletions, we first locate their positions based on the ID. If the data is present in the cache list, we remove it directly. Otherwise, i.e., it is stored in the index, we mark the corresponding position in the $T\_list$ for later deletion. When the size of the cache list exceeds a given limit, we efficiently reconstruct the entire index using the objects in {both the index and} the cache list, and then clear the cache. 
As depicted earlier, our construction strategy fully leverages the processing power of the GPU, enabling the index {\sf GTS} to be efficiently reconstructed. For instance, the index for 10 million objects can be rebuilt within just 2 seconds, minimizing disruption to the user. Moreover, as the entire index is reconstructed, update operations on {\sf GTS} have no adverse effects on search performance.

When answering similarity queries, we perform separate searches on both the index and the cache list, merging the results to retrieve final answers. Given that the cache list is much smaller than the dataset, we employ a brute-force table-based method for querying, utilizing GPU parallelism to compute the distance between each object in the cache list and the query and then validate all the results. Furthermore, our search strategy ensures accurate query outcomes by accounting for the presence of existing objects and newly inserted objects in the tree index and the cache list, while disregarding removed objects that are no longer accessible. 

\vspace{2mm}
\noindent
\textbf{Batch Updates.} Our index is stored in contiguous structures, which consist of a list-based structure and a distance table. However, updating operations on these structures can be costly, especially for list modifications. For instance, inserting elements in the middle of the list necessitates moving half of the elements, which is time-consuming. To address this issue, when significant bulk updates are encountered, we opt for direct index reconstruction. Leveraging our efficient parallel index construction method, we find that performing a direct batch reconstruction of the modified dataset is both feasible and effective, as demonstrated in {Section~\ref{sec:expcons}}.

\subsection{Complexity Analyses}
\label{sec:Complexity_Analysis}

\vspace{2mm}
\noindent
\textbf{Space Consumption of Index.} Our index consists of two components, i.e., the tree index $N\_list$ and the table list $T\_list$. Let $n$ denote the cardinality of the object set (i.e., $|O|$) and $N_c$ denote the node capacity. 
For the tree index, it is balanced and essentially consists of tree nodes stored consecutively in $N\_list$, so its space cost depends only on the number of tree nodes, which is $O(\frac{(N_c)^{h+1}-1}{N_c-1})$, i.e., $O(n)$, where $h$ is the height of the tree and equals \rev{$\lceil log_{N_c}{(|O|+1)} \rceil-1$}. 
Conversely, $T\_list$ records the partitioned data objects and their respective distances to the pivots. As the partitioning is applied to the entire dataset in a disjoint manner, the space cost of $T\_list$ directly correlates to the dataset size, yielding $O(n)$. Consequently, the overall space complexity of our index is $O(n)$.

\vspace{2mm}
\noindent
\textbf{Time Complexity of Index Construction.} Index construction {at each layer} primarily involves two key steps: pivot mapping and object partitioning. In the pivot mapping step at each level, we compute the distance between each data object and its corresponding pivot, {incurring the time complexity of $O(\frac{n}{C})$},
where $C$ represents the GPU concurrent computing power. For objects partitioning, the dataset is firstly sorted in ascending order based on the distances to the pivots, {incurring the cost of $O(\lceil \frac{n}{C} \rceil{\log^2n})$~\cite{tbd/JohnsonDJ21}. Next, the data partitioning with the complexity of $O(\frac{n}{C})$ is performed.} 
{In summary, the total overhead, which involves the index construction steps for $\log n$ layers, has the time complexity of $O(\lceil \frac{n}{C} \rceil {\log^3n})$.} Assuming the GPU computing power is on the same order of magnitude as the size of the dataset, the time complexity simplifies to $O(\log^3n)$. {For instance, recent GPU can be equipped with more than 16,000 cores and support 9$\times 10^{13}$ floating-point operations per second~\cite{urlgpu}.}

\vspace{2mm}
\noindent
\textbf{Time Complexity of Index Updating.} Index updating involves two main scenarios: batch updates and stream data updates. For batch updates, we perform a complete reconstruction of the index, so the time complexity is consistent with that of index construction. For stream data updates, we consider data insertion, data deletion, and index reconstruction when the cache list reaches its capacity. For data insertion, we directly store the new data in the cache list, resulting in a time complexity of $O(1)$. For data deletion, we locate the data based on its ID and then either mark it for deletion in the original index structure's $T\_list$ or immediately remove it from the cache list, resulting in a time complexity of $O(1)$. When the cache list is full, we execute index reconstruction, with time complexity consistent with that of index construction.
\section{Similarity Search}
\label{sec:Search}
In this section, we propose efficient GPU-based algorithms for similarity search in metric spaces using {\sf GTS}, including metric range query and metric $k$ nearest neighbor query.

\subsection{\rev{Metric Range Query in Batch}}

\begin{figure}[t]
    \centering
    \includegraphics[width=0.95\linewidth]{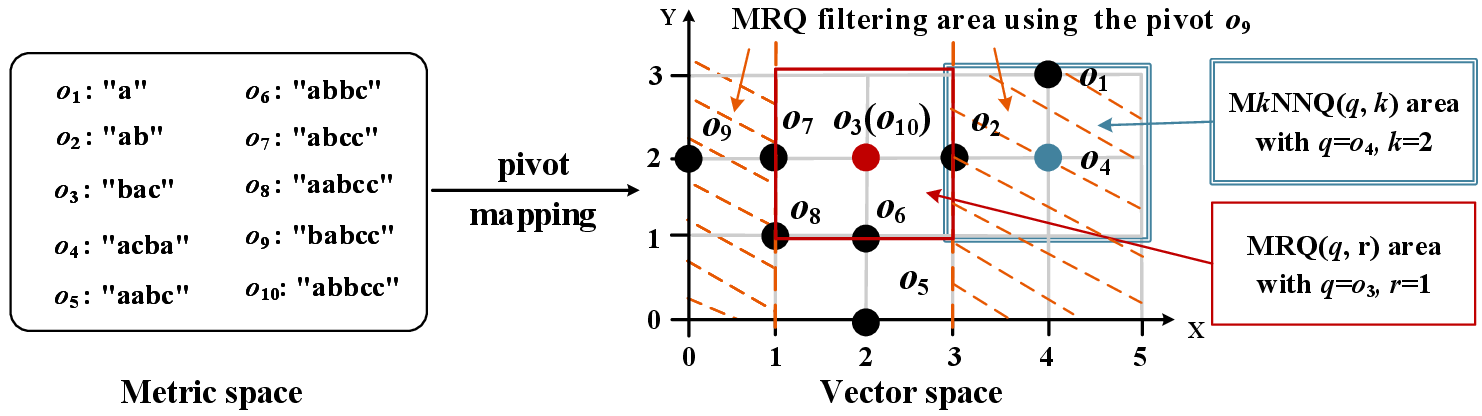}
    \vspace{-3mm}
    \caption{{Illustration of the filtering process}}
    \vspace{-3mm}
    \label{fig:filtering}
\end{figure}

Given a metric object set $O$, a range query with radius $r$ aims to find objects within $O$ whose distances to the query object $q$ are bounded by $r$. However, computing distances between the query object and unnecessary objects for answering range queries can be time-consuming. To accelerate metric range queries, we utilize the triangle inequality, following existing works~\cite{chen2020indexing}, to filter and validate objects without unnecessary distance computations.

\begin{lemma} \label{lemmafil}
\small
Given a pivot $p$, a query object $q$, and a search radius $r$ in a metric space, an object $o$ can be pruned if $\vert d(o,p) - d(q,p)\vert > r$.
\end{lemma}

The proof of Lemma~\ref{lemmafil} directly derives from the triangle inequality and is omitted here. To illustrate this, let's consider the example shown in Fig.~\ref{fig:filtering} for the metric space depicted in Fig.~\ref{fig:m-examp}. In this example, the query is $o_3$, and the search radius is 1. Applying Lemma~\ref{lemmafil} with pivot $o_9$, we can filter all the objects $o$ with $\vert d(o,o_9) - 2\vert > 1$ along X dimension, i.e., objects $o_1$, $o_4$, and $o_9$ that fall within the areas shaded by dashed lines in Fig.~\ref{fig:filtering} can be pruned. 

{The query process in GPU-based solutions starts with loading queries from CPU to GPU, processing the queries in GPU, and finally returning the results to CPU.} \rev{Similar to existing CPU-based metric indexes, {\sf GTS} adopts a top-down approach to answer similarity queries using triangle inequality-based pruning techniques. In the case of range queries, {\sf GTS} begins from the root node and systematically examines the tree nodes layer by layer, ensuring precise answers through the correctness of triangle inequality-based pruning, pruning only unqualified objects. For instance, given a query object $q$ and a tree node $N$, let $min\_dis$ be the minimum distance 
from the pivot $p$ of $N$ to the objects maintained by $N$. If $d(q,p)+r<min\_dis$, $N$ can be safely pruned by Lemma~\ref{lemmafil}. Otherwise, {\sf GTS} examines the subtrees of $N$ in the next layer.} However, a potential issue of exceeding the GPU's memory limit due to the space occupied by intermediate results from various queries can lead to a memory deadlock, where memory cannot be cleared to make room for further computations. 
To overcome this challenge, we develop a hierarchical two-stage query strategy that effectively avoids memory deadlocks. Specifically, we perform an iterative search of the index in a top-down manner to answer multiple queries \rev{in batches}, utilizing a table $Q\_Res$ to store intermediate results. Each element $E=\{N,q,r\}\in Q\_Res$ denotes that the node $N$ needs to be searched for the range query MRQ($q$, $r$).
To maximize the {utilization of} computing resources, {we manage the intermediate results of the query process by setting} an intermediate result size limit $size\_limit$ at layer $i$ to $\frac{size\_GPU}{h*N_c}$, where $size\_GPU$ is the GPU’s available memory size, $h$ is the tree height, and $N_c$ is the node capacity. Hence, the size of the intermediate answer at level $i+1$ ($i+1<h$) is bounded to $size\_limit*N_c=\frac{size\_GPU}{h}$. When the space consumption of $Q\_Res$ exceeds $size\_limit$, the queries are divided into various groups, and each group is processed linearly, ensuring the space consumption of the search process during the intermediate layers remains below the GPU memory limit. Note that the nodes in the bottom level $h$ can directly answer each query without incurring any extra cost. {Therefore, our proposed strategy effectively manages the memory size during the search of each level, preventing memory deadlocks and optimizing the utilization of computing resources.}

\begin{algorithm}[t]
\small
\SetNlSty{small}{}{:}
\LinesNumbered
\setstretch{1}
\DontPrintSemicolon
\caption{{Metric Range Query}}
\label{algo:range_query}
\KwIn{a set of range queries $Q$, the {\sf GTS} index}
\KwOut{the answers to the metric range queries}    
    initialize $Q\_Res$ to $\{\varnothing\}$\;
    \ForEach{$q_i\in Q$  \textbf{in parallel}}{
        $Q\_Res[i]\gets (N\_list[1],q_i,r_i)$\;    
    }       
    $N\_set\gets \{N\_list[1]\}$, $layer\gets 1$\;    
    $N_c\gets$ the node capacity of {\sf GTS}\;
    $h\gets$ the tree height of {\sf GTS}\; 
    Range\_Q($Q,Q\_Res,N\_set,layer,N_c,h$)\;    
    \SetKwFunction{moc}{Range$\_$Q}
    \textbf{Function} \moc{$Q,Q\_Res,N\_set,layer,N_c,h$}{
    \;    
        $size\_GPU\gets$ the available memory size of GPU\;
        $size\_limit\gets \frac{size\_GPU}{(h-layer+1)*N_c}$ \;
        $Q\_group\gets {Q}$ \;
        \If{the size of $Q\_Res$ >$size\_limit$}{            
            divide the $Q\_group$ according to $Q\_Res$ in parallel\;
        }
        \ForEach{ $Q'\in Q\_group$}{  
            initialize $Q'\_Res$ to $\{\varnothing\}$\;
            \ForEach{$E_i=\{N,q,r\}\in Q\_Res,j\in[0, N_c)$ \textbf{in parallel}}
            {    
               $N'\gets N\_list[N.pos+j]$ \;
                \If{$N'$ cannot be pruned for MRQ($q,r$) by Lemma~\ref{lemmafil}}
                {                    
                    $Q'\_Res[i*N_c+j]\gets \{N\_list[N.pos+j],q,r\}$\;                    
                }                   
                               
            }            
    $N'\_set\gets$ the nodes in $Q'\_Res$\;
            \textbf{if} $layer=h$ \textbf{then}  output the answers to the queries of $Q'$ by verifying the objects in the nodes of $N\_set'$\;
            \textbf{else}  Range\_Q($Q',Q'\_Res,N'\_set,layer+1,N_c,h$)\;            
        }     
  }
\end{algorithm}

Algorithm~\ref{algo:range_query} outlines the process of the two-stage \rev{metric range query}. This algorithm takes as input a set of range queries $Q$ and the {\sf GTS} index, and outputs the answer to each range query. Firstly, we initialize all the elements in the intermediate result table $Q\_Res$, and set $N\_set$ to the root node and the current search layer $layer$ to 1.
Additionally, we obtain the node capacity $N_c$ and the tree height $h$ from the {\sf GTS} index (lines 1--6). Next, the algorithm invokes $Range\_Q$ function to answer the range queries (line 8).  It initially computes the size limit based on GPU's available memory size $size\_GPU$ and partitions the queries if needed (lines 9-10). Next, for each query group $Q'$, the function first initializes a new intermediate result table $Q'\_Res$ for the next layer {(line 15)}. It then concurrently finds the node $N$ that cannot be pruned by the range query $MRQ(q,r)$ via Lemma~\ref{lemmafil}, and updates the corresponding intermediate result $Q'\_Res$ {(lines 16--20)}. Thereafter, the function collects all the nodes $N'\_set$ in $Q'\_Res$ {(line 22)}. Finally, if the current layer is at the bottom level, the function uses the nodes in $N'\_set$ to answer each range query (line 22). Otherwise, it proceeds to search the next layer to answer the range queries (line 23).

\subsection{\rev{Metric \textit{k}NN Query in Batch}}
A metric $k$ nearest neighbor query (M\textit{k}NNQ) aims to find $k$ objects in $O$ that are most similar to $q$. Consider a metric $k$ nearest neighbor query instance depicted in Fig.~\ref{fig:filtering}. The query object $q$ is $o_3$, and $k$ is 2. The answer to the query M\textit{k}NNQ($q=o_3,k=1$) is $\{o_3,o_{10}\}$. To accelerate the search process, we can apply the following lemma. 

\begin{lemma} \label{lemmafilk}
\small
Given a pivot $p$, a query object $q$, and an integer $k$ in a metric space, assuming that in the current search stage, the distance between $q$ and its current $k$-th nearest neighbor is $d(q,k_{cur})$, then an object $o$ can be pruned if $\vert d(o,p) - d(q,p)\vert \geq d(q,k_{cur})$.
\end{lemma}

The proof of Lemma~\ref{lemmafilk} is straightforward. If $\vert d(o,p) - d(q,p)\vert \geq d(q,k_{cur})$, it indicates that there exist at least $k$ objects whose distances to $q$ are not greater than $o$'s distance. Therefore, $o$ can be safely pruned. Let's apply Lemma~\ref{lemmafilk} to the example query M\textit{k}NNQ($o_4$, 2) shown in Fig.~\ref{fig:filtering}. 
During the search, if we have already visited objects $o_5$ and $o_6$ and obtained the distance boundary $d(q,k_{cur})=2$, then object $o_7$ can be safely pruned using the pivot $o_9$ as $|d(o_7,o_9)-d(o_4,o_9)|=3>2$.

\rev{Two typical solutions exist for $k$NN queries, including the best-first strategy and the range query-based strategy~\cite{ChavezPromximity,chen2020indexing}. Existing GPU-based indexes mainly apply the best-first strategy, which cannot be parallelized due to its iterative process relying on visited nodes to prune unvisited nodes, thereby suffering from unbalanced workload and memory issues~\cite{sc/GaihreZWLSDLL21}. To address this, we propose an alternative approach that capitalizes on GPUs' capacity for simultaneous computations by concurrently traversing tree nodes in a top-down and by-level manner. It progressively uses the distance boundary, initially set to infinity, and selects the current best $k$ objects to derive the narrowed distance boundary, subsequently pruning lower-level tree nodes using Lemma~\ref{lemmafilk}.} 

Algorithm~\ref{algo:knn_query} presents the process of the \rev{metric $k$NN query in batch}. Similar to the metric range query algorithm, the metric $k$NN query algorithm initializes the intermediate result table $Q\_Res$, the set of nodes $N\_set$ in the current search layer $layer$, the node capacity $N_c$, and the tree height $h$ (line 1). It then invokes the $Knn\_Q$ function to answer the queries (line 2). The $Knn\_Q$ function first computes the query groups $Q\_group$ in the same manner as Algorithm~\ref{algo:range_query} (line 4). Next, for each group of queries $Q'$, the function initializes the elements in the intermediate result table $Q'\_res$ (line 6). After computing the distance from each child node $N'$ of $N$ to the query object $q$, where $\{N,q\}\in Q\_Res$ (lines 7--11), the function sorts those distances via parallelized encoding and sorting strategies as in Algorithm~\ref{algo:objects_partitioning} (line 12). This ensures that the distances from the objects of different nodes to the same query objects are located in contiguous segments in the temp table $T\_tmp$. Thereafter, the function locates the $k$-th maximum distance $dis\_k[j]$ of each query M\textit{k}NNQ($q,k$) via $T\_tmp$ (line 13) and filters the unnecessary nodes by Lemma~\ref{lemmafilk} (lines 14--17). Finally, the function decides whether to answer each M\textit{k}NNQ query based on the current layer (line 19) or search the next layer (line 20).

\begin{algorithm}[t]
\small
\SetNlSty{small}{}{:}
\LinesNumbered
\setstretch{1}
\DontPrintSemicolon
\caption{Metric \textit{k}NN Query}
\label{algo:knn_query}
\KwIn{a set of M$k$NN queries $Q$, the {\sf GTS} index}
\KwOut{the answers to the metric $k$NN queries}    
    initialize the $Q\_Res$, $N\_set$, $N_c$, $layer$, and $h$ in the same step of Metric Range Query algorithm \;    
    Knn$\_$Q($Q,Q\_Res,N\_set,layer,N_c,h$)\;  
    \SetKwFunction{moc}{Knn\_Q}
    \textbf{Function} \moc{$Q,Q\_Res,N\_set,layer,N_c,h$}{
    \;    
        compute the $Q\_group$ according to  $Q$ as Algo.~\ref{algo:range_query} lines 9--13\;
        \ForEach{ $Q'\in Q\_group$}{  
           {initialize $Q'\_Res$ to $\{\varnothing\}$} \;
            \ForEach{{{$E_i=\{N,q,k,d\}\in Q\_Res,j\in[0,N_c)$}\textbf{in parallel}}}{   
                $N'\gets N\_list[N.pos+j]$ \;   
                $dis\gets$ the distance between $N'.pivot$ and $q$\;
                $Q'\_Res[i*N_c+j]\gets \{N',q,k,dis)$\}\;                   
            }                            
            $T\_tmp\gets$ the sorted distances in $Q'\_Res$ via parallelized encoding and sorting strategies as Algorithm~\ref{algo:objects_partitioning}\;
            locate the current $k$-th maximum distance $dis\_k[i]$ of each query M\textit{k}NNQ($q_i,k_i$) via $T\_tmp$\;
            \ForEach{$E_i=\{N,q,k,d\}\in Q'\_Res$ \textbf{in parallel}}{ 
                $dis\gets$ the distance stored in $Q'\_Res[i]$\;
                \If {$N$ can be pruned by Lemma~\ref{lemmafilk} using $dis\_k$} 
                {
                $Q'\_Res[i]\gets \varnothing$\;
            }
            }            
            $N'\_set\gets$ the nodes in $Q'\_Res$\;
            \textbf{if} $layer=h$ \textbf{then} output the answers to the queries in $Q'$ by verifying  the objects in the nodes of $N'\_set$\;     
            \textbf{else} Knn\_Q($Q',Q'\_Res,N'\_set,layer+1,N_c,h$)\;   
            
        }        
  }
\end{algorithm}

\vspace{0.2cm}
\noindent
\textbf{Remark.} \rev{Notably, the proposed method {\sf GTS} holds the potential to handle multi-column scenarios. For instance, within the established {\sf PM-Tree} framework~\cite{icde/FranzkeEZR16}, we can create a {\sf GTS} index for each column and address multi-column queries by progressively combining the results of each queried attribute using Fagin’s algorithm ~\cite{jcss/FaginLN03} and pigeon-hole principle~\cite{pvldb/Zhu22}. }

\subsection{Cost Model}
\label{sec:cost}
In the following, we present a cost model for the similarity search, including {MRQ} and {M\textit{k}NNQ}. As the queries are answered in parallel, following the GPU's parallelized schedule, our focus here is to estimate the cost for each single query.

We begin by analyzing the time cost of the metric range query. The proposed search process for the metric range query {follows a top-down approach, traversing \rev{$\log_{N_c} {n}$} layers in the {\sf GTS} index}. Thus, the time complexity of the range query is \rev{$O(\sum_{i=1}^{\log_{N_c} {n}} \lceil \frac{S_i}{C}\rceil \cdot$ $\log^2 S_i)$}, where $N_c$ represents the node capacity, $S_i$ denotes the intermediate result size at layer $i$, and $C$ indicates the GPU concurrent computing power. Now, let's evaluate the size of $S_i$. We start by assessing the probability that an object cannot be pruned by the index. Objects are partitioned into different subsets by pivot mapping at each level, and the distances between an object $o$ and the pivots can be treated as random variables $X_1,X_2,\cdots$. According to Lemma~\ref{lemmafil}, an object $o$ in node $N$ with pivot $p$ cannot be pruned if $|d(o,p)-d(q,p)|\leq r$. Notably, an object cannot be excluded only if it cannot be pruned by pivots at every level. Thus, the probability that an object cannot be pruned at level $i$ can be computed as:
\begin{equation}
\small
P_r(o\ is\ not\ pruned)=P_r(|X-Y|\leq r)^{i}.
\end{equation}
Here, $Y$ denotes the distance from the query $q$ to the pivot $p$, which follows the same distribution of $X$ since $q$ can also be regarded as a random object. 
Since $X$ and $Y$ are two independent, identically distributed random variables with variance $\sigma^2$, the mean and variance of ($X-Y$) are 0 and $2\sigma^2$, respectively. Applying Chebyschev’s inequality\cite{ChavezPromximity}, we have
\begin{equation}
\label{equ:prunprob}
\small
P_r(|X-Y|\leq r)^{i}\geq(1-\frac{2\sigma^2}{r^2})^{i}.
\end{equation}

Based on the above analysis, we can estimate the lower bound time complexity of MRQ as \rev{$O(\sum_{i=1}^{\log_{N_c} n} i^2  \lceil \frac{N_c^i(1-\frac{2\sigma^2}{r^2})^{i}}{C}\rceil \log^2 N_c)$}, where $n$ is the size of objects and $N_c$ denotes the node capacity. Recall that {M\textit{k}NNQ} can be answered by MRQ, and its search process also follows the top-down search manner of MRQ. Therefore, the above time complexity of MRQ also applies to {M\textit{k}NNQ}.

\noindent
\textbf{Discussions.} The time complexity analysis above highlights the strong correlation between search efficiency and two key factors: the node capacity $N_c$ and the dataset characteristics. 
A larger $N_c$ can lead to a reduction in the tree height, resulting in lower overhead to search different layers. However, it also allows fewer pivots to be used for pruning, limiting the pruning capability, as described in Equation~\eqref{equ:prunprob}.  
Let's consider the three situations as below:

{\em (1) $n\ll C$}: In this scenario, the GPU computing power exceeds the dataset size, and the overhead becomes  \rev{$O(\sum_{i=1}^{\log_{N_c} {n}} i^2 \log^2 N_c)$}, i.e., \rev{$O((\log_{N_c} {n})^3\cdot \log^2N_c)=O(\frac{\log^3n}{\log N_c})$}. To optimize the efficiency, a large $N_c$ is preferred, as it reduces the tree height, hence lowering the overhead to search different layers.

{\em (2) $n\gg C$}: In situations with large data sizes or high query throughput, the GPU cannot handle all the computing tasks simultaneously. The time complexity is simplified to 
\rev{$O(n (1-(\frac{2\sigma^2}{r^2}))^{\log_{N_c} n}$ $\log^2n)$}. Here, a smaller $N_c$ results in higher efficiency.

{\em (3) $O(n)=O(C)$}: The search cost becomes complex and cannot be analyzed directly. However, considering that a few pivots can significantly reduce unnecessary distance computations~\cite{jda/MaoMM12,pvldb/ChenGZJYY17}, it is suggested to use a relatively small number of pivots (hence, lower tree height) to strike a balance between the GPU parallelism and objects pruning capability, achieving high \rev{search efficiency}.

\begin{table}
\caption{{Dataset statistics}}
\label{tab:datasets}
\small
\vspace{-0.3cm}
\renewcommand\arraystretch{1}
\begin{tabular}{|p{2cm}<{\centering}|p{3cm}<{\centering}|p{3cm}<{\centering}|p{4.1cm}<{\centering}|}
\hline
\textbf{Dataset} & \textbf{Cardinality} & \textbf{Dimensionality} & \textbf{Distance Metric}\\
\hline 
\textit{Words} & $611,756$ & 1$\sim$34  & \textit{Edit distance}   \\ \hline
\textit{T-Loc} & $10,000,000$ & $2$  & $L_{2}$-norm \\ \hline
\textit{Vector} & $200,000$ & $300$  & \textit{Word cosine distance} \\ \hline
\rev{\textit{DNA}} & \rev{$1,000,000$} & \rev{$108$}  & \rev{\textit{Edit distance}} \\ \hline
\textit{Color} & \rev{$5,000,000$} & $282$  & $L_{1}$-norm  \\ \hline
\end{tabular}
\vspace{-0.4cm}
\end{table}

\begin{table}
\caption{{Evaluation parameters}}
\small
\vspace{-0.3cm}
\renewcommand\arraystretch{0.95}
\label{tab:quepara}
\setlength{\tabcolsep}{1.15mm}{
\begin{tabular}{| p{6.5cm}<{\centering}  | p{6.5cm}<{\centering} |} \hline
\textbf{Parameter} & \textbf{Value} \\ \hline
Search radius \textit{r} (x0.01\%)& 1, 2, 4, \textbf{ 8},  16, 32   \\ \hline
Integer \textit{k} & 1, 2, 4, \textbf{8}, 16, 32\\ \hline
Node capacity $N_c$ & 10, \textbf{20}, 40, 80, 160, 320 \\ \hline
{Number of queries in a batch} & 16, 32, 64,  {\textbf{128}, 256, 512}   \\ \hline
\rev{distinct data proportion (\%)} & \rev{20, 40, 60, 80, \textbf{100}}   \\ \hline
\textit{Cardinality} (\%) & \textbf{20}, 40, 60, 80, \textbf{100} \\ \hline
\end{tabular}
}
\vspace{-0.4cm}
\end{table}

\begin{table*}[t]
\small
\caption{\rev{Index construction cost of different methods}}
\label{tab:construction_cost}
\vspace{-0.2cm}
\renewcommand\arraystretch{1}
\setlength{\tabcolsep}{1.12mm}{
\begin{tabular}{|p{1.7cm}<{\centering}|p{0.88cm}<{\centering} |p{1.05cm}<{\centering} |p{0.88cm}<{\centering} |p{1.05cm}<{\centering} |p{0.88cm}<{\centering} |p{1.05cm}<{\centering} |p{0.88cm}<{\centering} |p{1.05cm}<{\centering} |p{0.88cm}<{\centering} |p{1.05cm}<{\centering} |} \hline

 \multirow{3}{*}{Method}& \multicolumn{2}{c|}{\textit{Words}}  & \multicolumn{2}{c|}{\textit{T-Loc}}& \multicolumn{2}{c|}{\textit{Vector} }& \multicolumn{2}{c|}{\textit{\rev{DNA}}}& \multicolumn{2}{c|}{\textit{Color}} \\ \cline{2-11}
& Time &  \rev{Storage} & Time &   \rev{Storage} & Time &   \rev{Storage} & Time &   \rev{Storage} & Time &   \rev{Storage} \\ 
\specialrule{0em}{-0.5pt}{-0.5pt}
& (s) &  \rev{(MB)} & (s) & \rev{(MB)}& (s)& \rev{(MB)}& (s)&  \rev{(MB)}& (s) & \rev{(MB)}\\ \hline
{\sf BST}  & 12.96 & 2.34 & 5.76 & 38.15 & 1.24 & 0.77 & 635.29 & 3.82 & 4.33 & 3.82  \\ \hline
{\sf EGNAT} & 89.63 & 430 &   /   &   /   & 141.87 & 474.99 & 3447.89 & 637 & 676.01 & 2137.72  \\ \hline
{\sf MVPT} & 0.41 & 2.52 & 10.17 & 38.099 & 12.17 & 0.76 & 33.02 & 3.85 & 35.81 & 3.81  \\ \hline
{\sf GPU-Tree} & 3.94 & 6.2 & 87.44 & 105.06 & 1.21 & 2.09 & 28.7 & 10.14 & 6.34 & 10.5  \\ \hline
{\sf LBPG-Tree} &   /   &   /   & 0.073 & 40.14 &   /   &   /   &   /   &   /   & 0.32 & 146.37  \\ \hline
{\sf GANNS} &   /   &   /   &   /   &   /   & 2.35 & 48.82 &   /   &   /   & 10.28 & 244.14  \\ \hline
{\sf GTS} & 0.11 & 2.63 & 1.1 & 41.1 & 0.2 & 1.05 & 2.47 & 4.11 & 0.54 & 4.1  \\ \hline
\end{tabular}
}
\vspace{-2mm}
\end{table*}

\begin{table}[t]
\small
\caption{\rev{Update time of {\sf GTS} under different cache table size}}
\label{tab:update_cost}
\vspace{-0.2cm}
\renewcommand\arraystretch{1}
\setlength{\tabcolsep}{1.12mm}{
\begin{tabular}{|p{2cm}<{\centering} |p{2.1cm}<{\centering} |p{2.1cm}<{\centering} |p{2.1cm}<{\centering} |p{2.1cm}<{\centering}|p{2.1cm}<{\centering} |} \hline
Dataset & {0.01}KB  & 0.1KB & 1KB & 5KB & 10KB \\ \hline
\textit{Words}~\cite{urlmoby} & 0.01729s & 0.01403s & \textbf{0.01368}s & 0.01389s & 0.01401s \\ \hline
\textit{T-Loc}~\cite{pvldb/GhoshACHSL18} & 0.06865s & 0.02144s & 0.01667s & 0.01561s & \textbf{0.01559}s \\ \hline
\textit{Vector}~\cite{bilbao2018automatic} & 0.02773s & 0.01948s & 0.01862s & \textbf{0.01852}s & 0.01876s \\ \hline
\textit{\rev{DNA}}~\cite{urlDNA} & 0.46009s & 0.35260s & 0.33381s & \textbf{0.31736}s & 0.31738s \\ \hline
\textit{Color}~\cite{bolettieri2009} & 0.06106s & 0.03200s & 0.02992s & \textbf{0.02251}s & 0.02254s \\ \hline
\end{tabular}
}
\vspace{-3mm}
\end{table}

\section{Experiments}
\label{sec:exp}
In this section, we conduct empirical experiments to evaluate the performance of our proposed GPU-based tree index {\sf GTS}, including the construction and update costs, the similarity search performance, and the scalability.

\subsection{Experimental Settings}
\label{sec:exp-sett}
\textbf{Datasets.} We use \rev{five} real-life datasets in our study: {\bf (i)} \textit{Words}~\cite{urlmoby} that contains proper nouns, acronyms, and compound words sourced from the Moby project, using edit distance to measure the similarity between words; {\bf (ii)} \textit{T-Loc}~\cite{pvldb/GhoshACHSL18} that contains geographical locations of 10$M$ Twitter-users, where the $L_{2}$-norm distance is employed as the distance metric; {\bf (iii)} \textit{Vector}~\cite{bilbao2018automatic} that includes 200K word embeddings of dimension 300 trained on the Spanish Billion Words Corpus, where the word cosine distance~\cite{urlwcd} is the distance metric; \rev{{\bf (iv)} \textit{DNA}~\cite{urlDNA} that consists of 1 million DNA data, where the edit distance is employed as the distance metric;} and {\bf (v)} \textit{Color}~\cite{bolettieri2009} that contains image features extracted from \textit{Frickr}, where the $L_{1}$-norm distance is employed to measure the similarity between features.
Table~\ref{tab:datasets} summarizes all the applied datasets.

\noindent
\textbf{Baselines.}
To evaluate our proposed index {\sf GTS}, we first conducted a comparison against: {\bf (i)} three most efficient CPU-based main-memory approaches for similarity search~\cite{chen2020indexing}, including {\sf BST}~\cite{tse/KalantariM83},  {\sf EGNAT}~\cite{iccS/MarinUB07,is/NavarroP11}, and {\sf MVPT}~\cite{sigmod/BozkayaO97,tods/BozkayaO99}; {\bf (ii)} two GPU-based baselines for general metric spaces, including the {\sf GPU-Table} that computes the distances between query and all the objects to answer MRQ and leverage Dr. Top-\textit{k} algorithm~\cite{sc/GaihreZWLSDLL21} to answer M\textit{k}NNQ, and the {\sf GPU-Tree} that implements the SOTA GPU-based tree index {\sf G-PICS}~\cite{tkde/LewisT22} strategy for general similarity search on single GPU by constructing multiple MVP-Trees; \rev{and {\bf (iii)} two SOTA GPU-based special-purpose solutions, including {\sf LBPG-Tree}~\cite{tsc/KimLC22} that constructs R-Trees on GPU for similarity search and GPU-based graph method {\sf GANNS}~\cite{icde/YuWZQZL22} for vector \textit{k}NN search.}

\noindent
\rev{\textbf{Remark.} Due to {\sf LBPG-Tree} supports similarity search only on vector data with $L_p$-norm distance on \textit{T-Loc} and \textit{Color}, and {\sf GANNS} is designed for vector similarity search on \textit{T-Loc}, \textit{Vector}, and \textit{Color}, this paper reports only the available results of these two special-purpose baselines.}

\noindent
\rev{ \textbf{Configuration.} All experiments were conducted on an Ubuntu server equipped with an Intel Core i9-10900X CPU, 128G of host memory, and an Nvidia Geforce RTX 2080 Ti GPU with 11G of device memory. The source code, data, and/or other artifacts have been made available~\cite{urlcode}.}

\noindent
\textbf{Parameters and Performance Metrics.} In this study, we evaluate the performance of our proposed method {\sf GTS} and its competitors, by exploring the impact of several key parameters. Specifically, we vary the search radius $r$ for MRQ, the $k$ for M\textit{k}NNQ, the node capacity that controls the number of child nodes each tree node has, the number of queries in a \rev{batch}, \rev{distinct data proportion}, and the cardinality (i.e., the percentage relative to the entire dataset).
Table~\ref{tab:quepara} details the settings of key parameters, with default values shown in bold. \rev{In terms of cardinality, the default size of \textit{Color} dataset is set to 20\% to ensure baseline methods are executable within the limited GPU memory, while the default value is 100\% for other datasets}.
To validate the efficiency and effectiveness of {\sf GTS}, we measure several metrics, \rev{including index construction cost, update time, and throughput}. Each measurement is based on the average performance across 100 randomly generated queries.

\begin{figure}[t]
\begin{center}
\subfigtopskip=-6pt
\subfigcapskip=-3pt
\includegraphics[height=0.4cm]{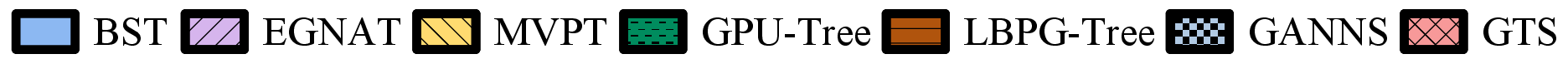}\vspace{0.16cm}

\subfigure[\textit{Streaming data updates}]{
  \label{fig:rnn-02}
  \includegraphics[width=0.6\textwidth]{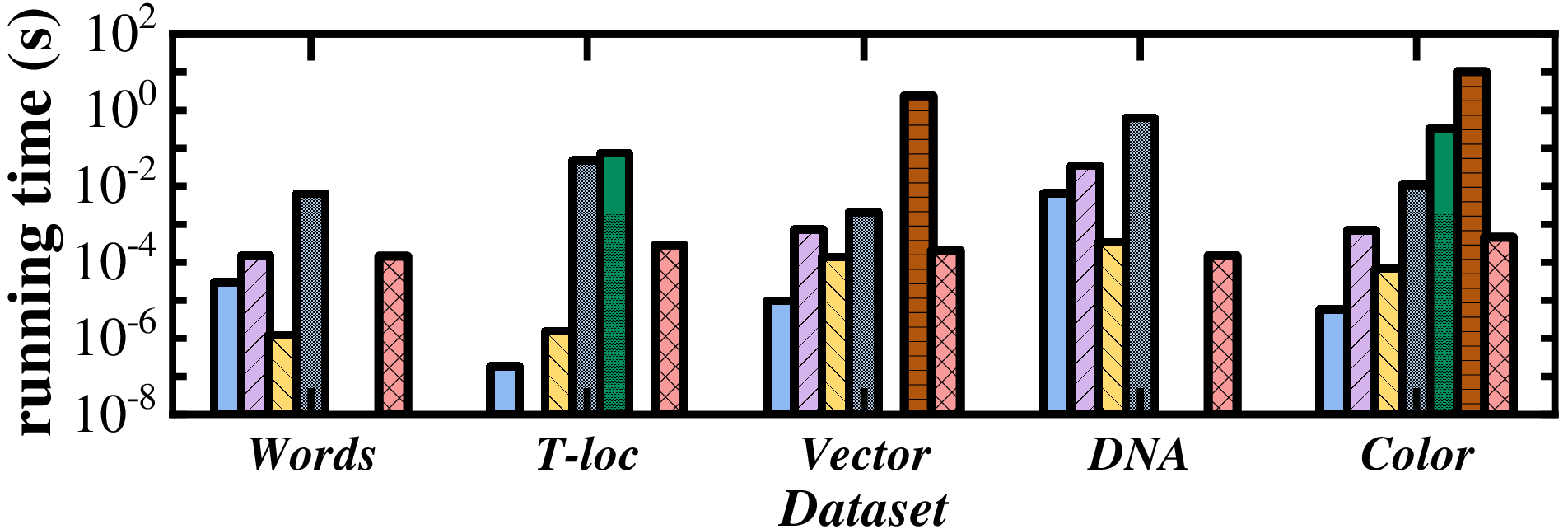}
}

\subfigure[\textit{Batch updates}]{
  \label{fig:rnn-02}
  \includegraphics[width=0.6\textwidth]{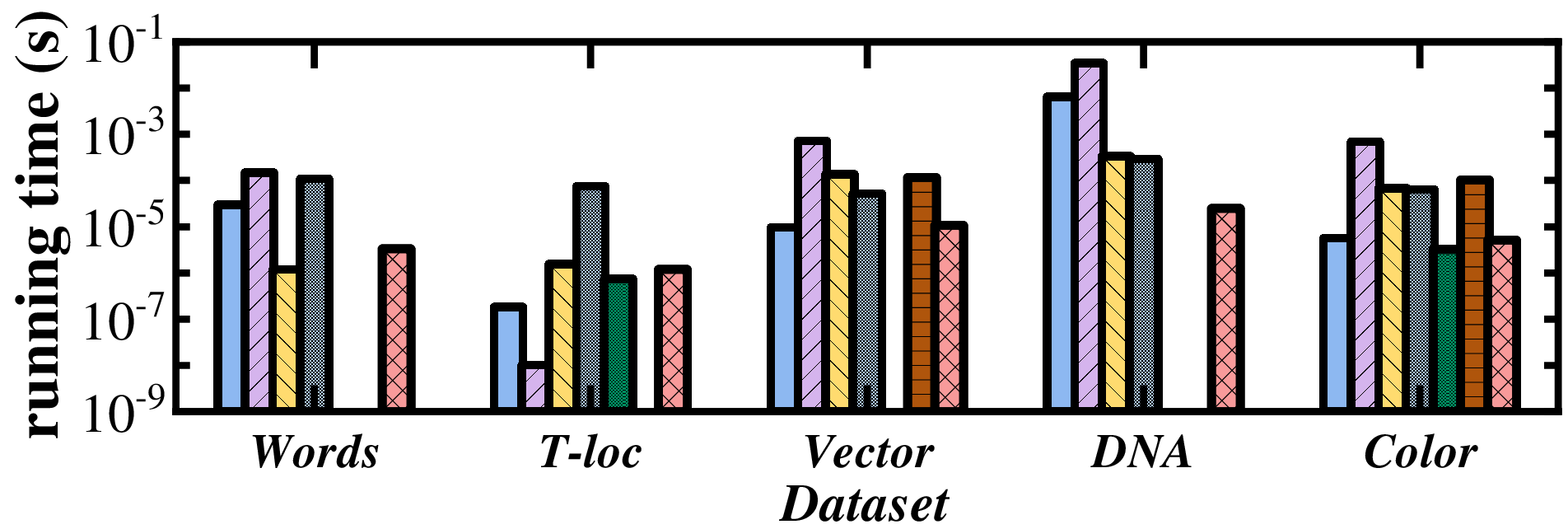}  
}
\vspace{-0.4cm}
\caption{\rev{Update cost}}
\vspace{-0.5cm}
\label{fig:exp-update}
\end{center}
\end{figure}

\begin{figure}[t]
\begin{center}
\subfigtopskip=-6pt
\subfigcapskip=-3pt
\subfigure[\textit{Words}]{
  \label{fig:rnn-02}
  \includegraphics[width=0.28\textwidth]{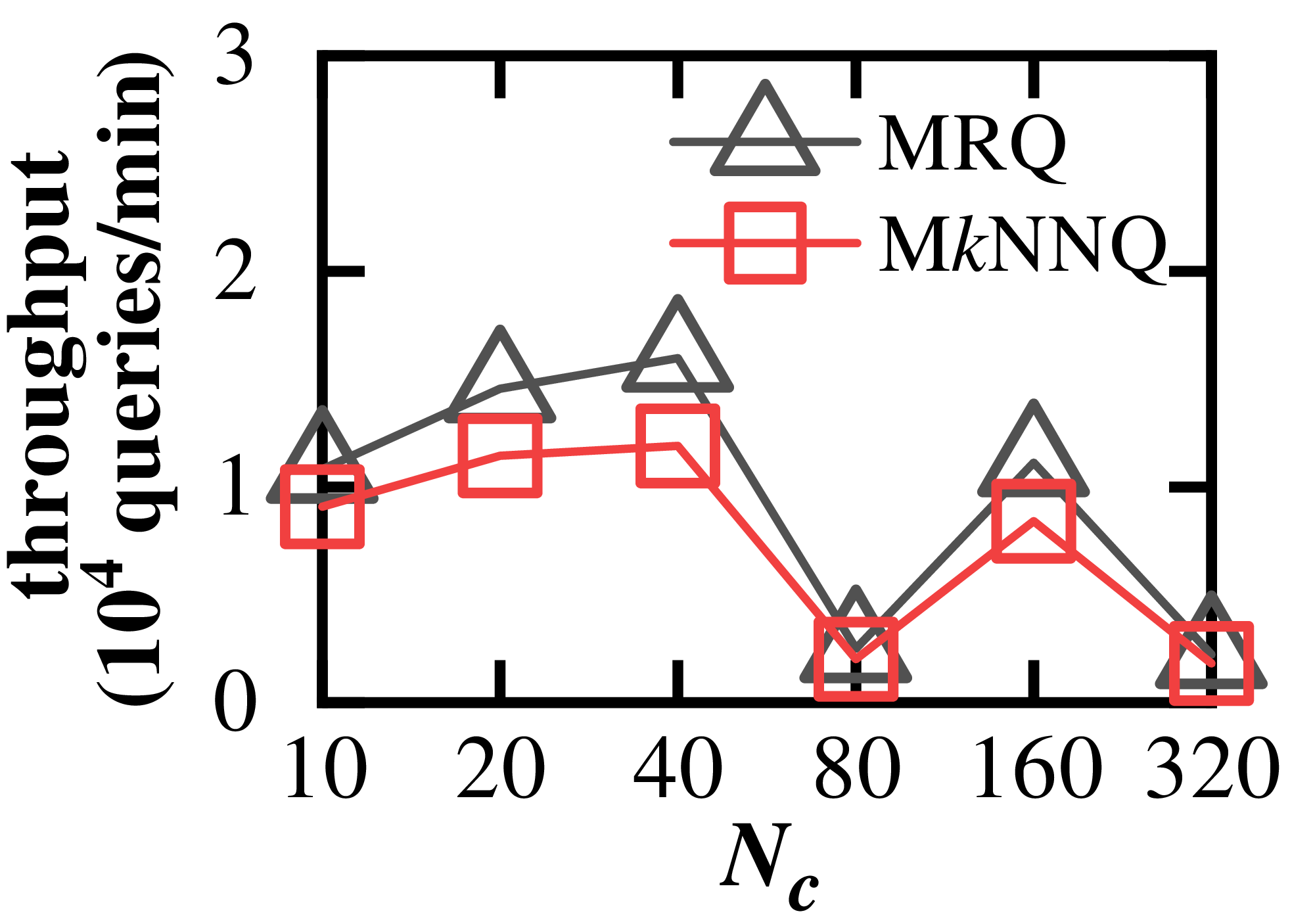}  
}
\subfigure[\textit{Color}]{
  \label{fig:rnn-02}
  \includegraphics[width=0.28\textwidth]{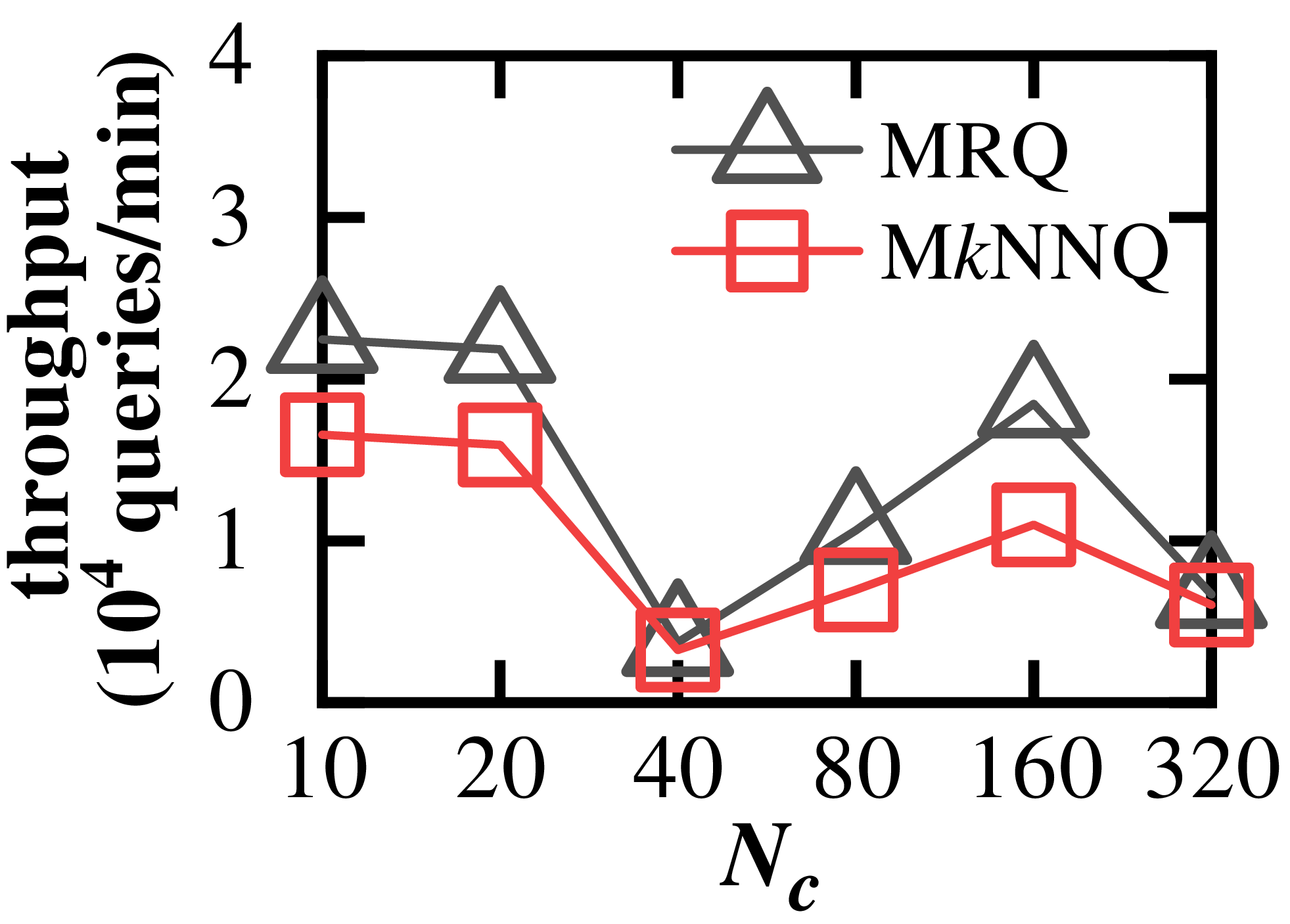}  
}
\vspace{-0.4cm}
\caption{\rev{Effect of the node capacity $N_c$}}
\vspace{-0.4cm}
\label{fig:exp-partition}
\end{center}
\end{figure}

\begin{figure*}[t]
\begin{center}
\subfigtopskip=-6pt
\subfigcapskip=-3pt
\includegraphics[height=0.4cm]{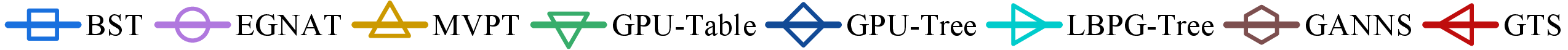}\vspace{0.15cm}

\subfigure[MRQ and \textit{Words}]{
  \label{fig:rnn-02}
  \includegraphics[width=0.23\textwidth]{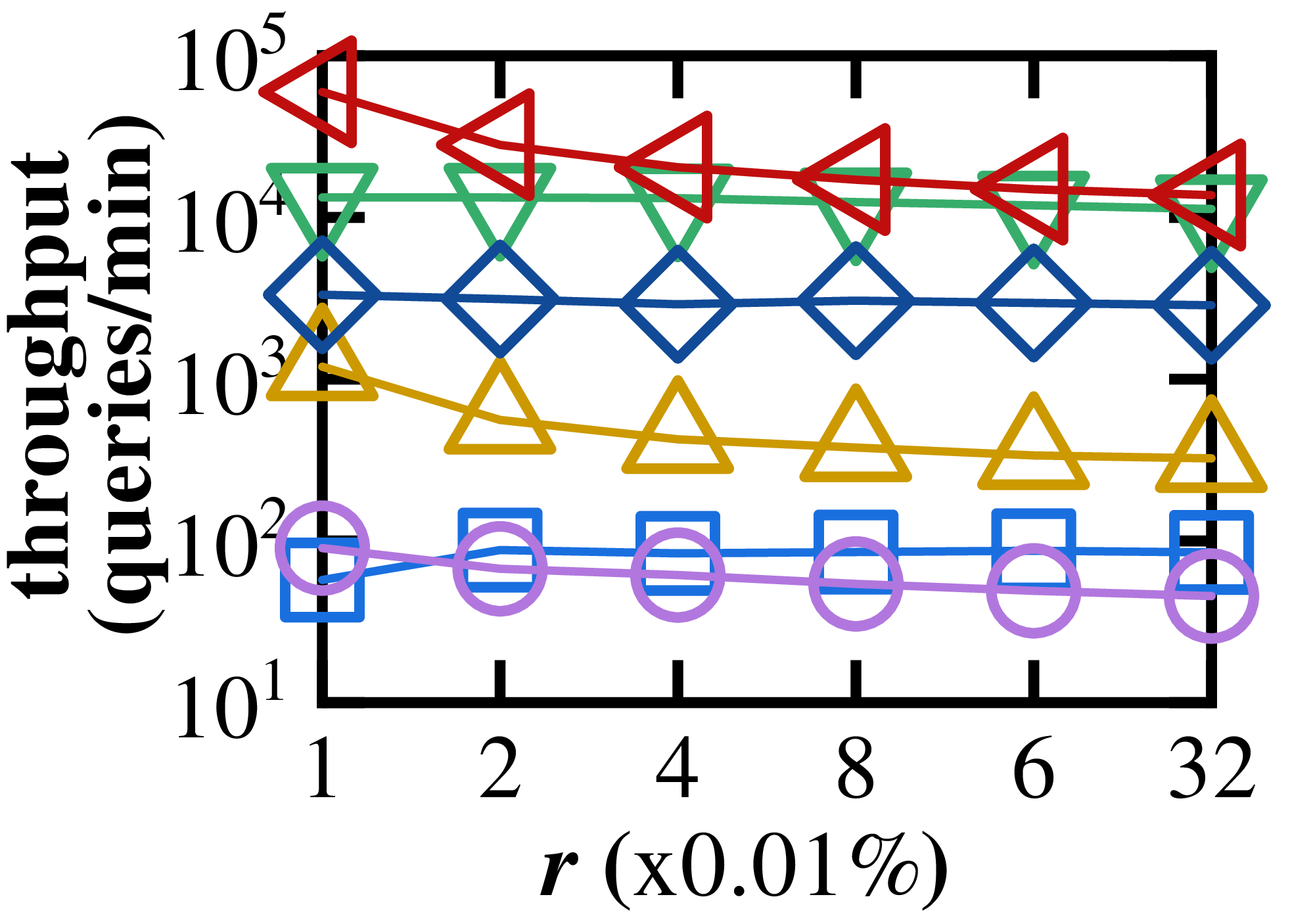}
}
\subfigure[MRQ and \textit{T-Loc}]{
  \label{fig:rnn-02}
  \includegraphics[width=0.23\textwidth]{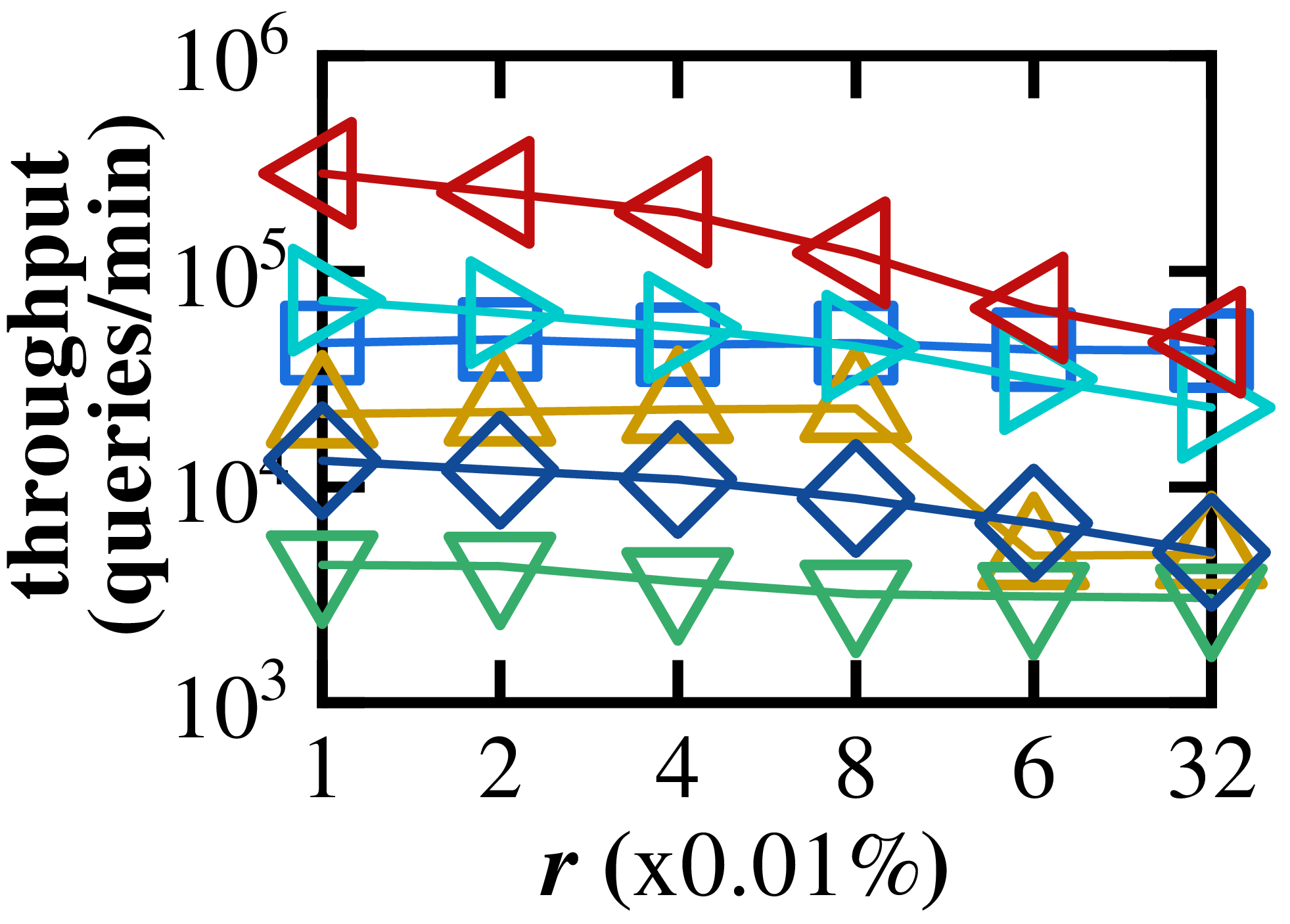}
}
\subfigure[MRQ and \textit{Vector}]{
  \label{fig:rnn-02}
  \includegraphics[width=0.23\textwidth]{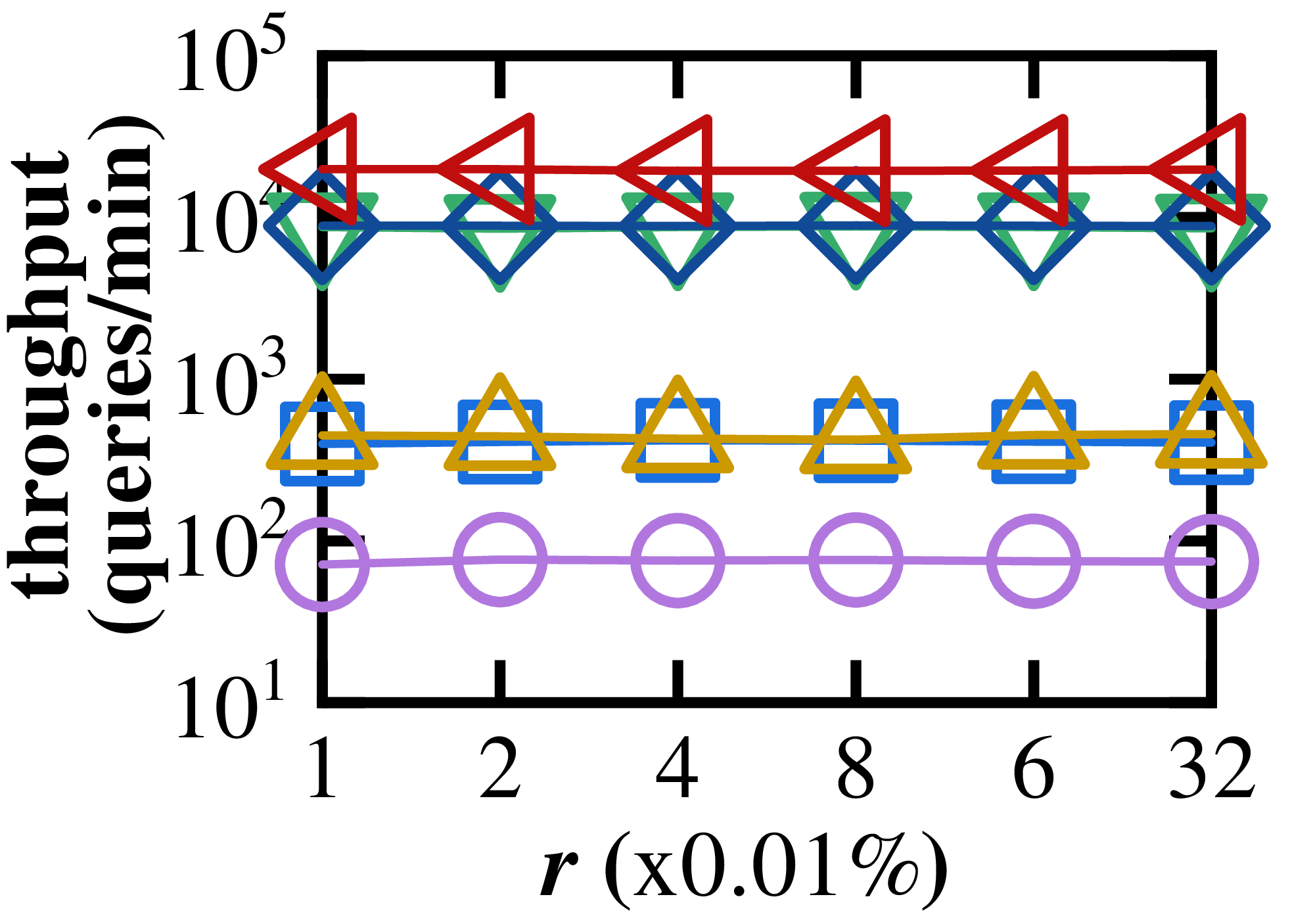}
}
\subfigure[MRQ and \rev{\textit{DNA}}]{
  \label{fig:rnn-02}
  \includegraphics[width=0.23\textwidth]{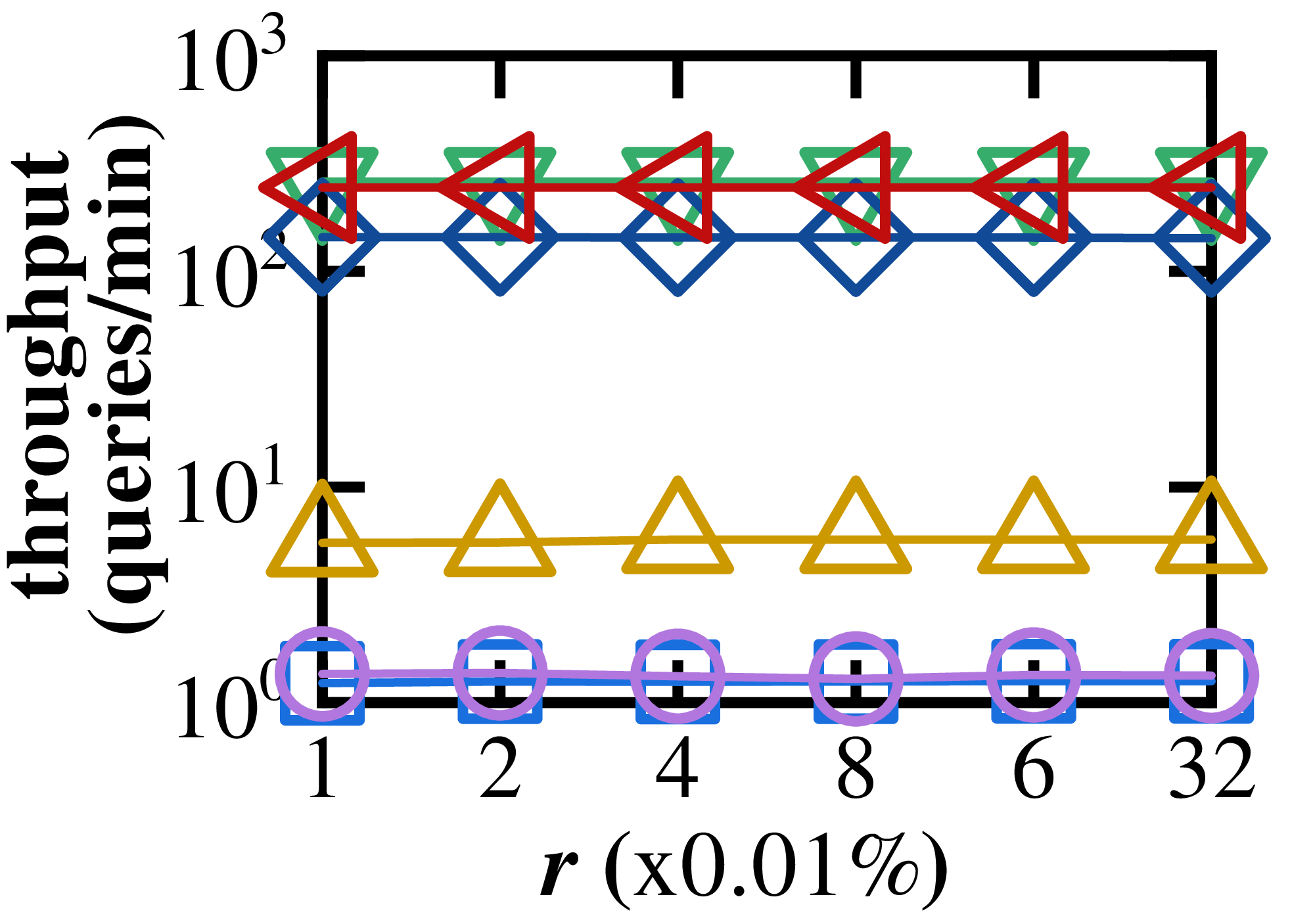}
}

\subfigure[MRQ and \textit{Color}]{
  \label{fig:rnn-02}
  \includegraphics[width=0.23\textwidth]{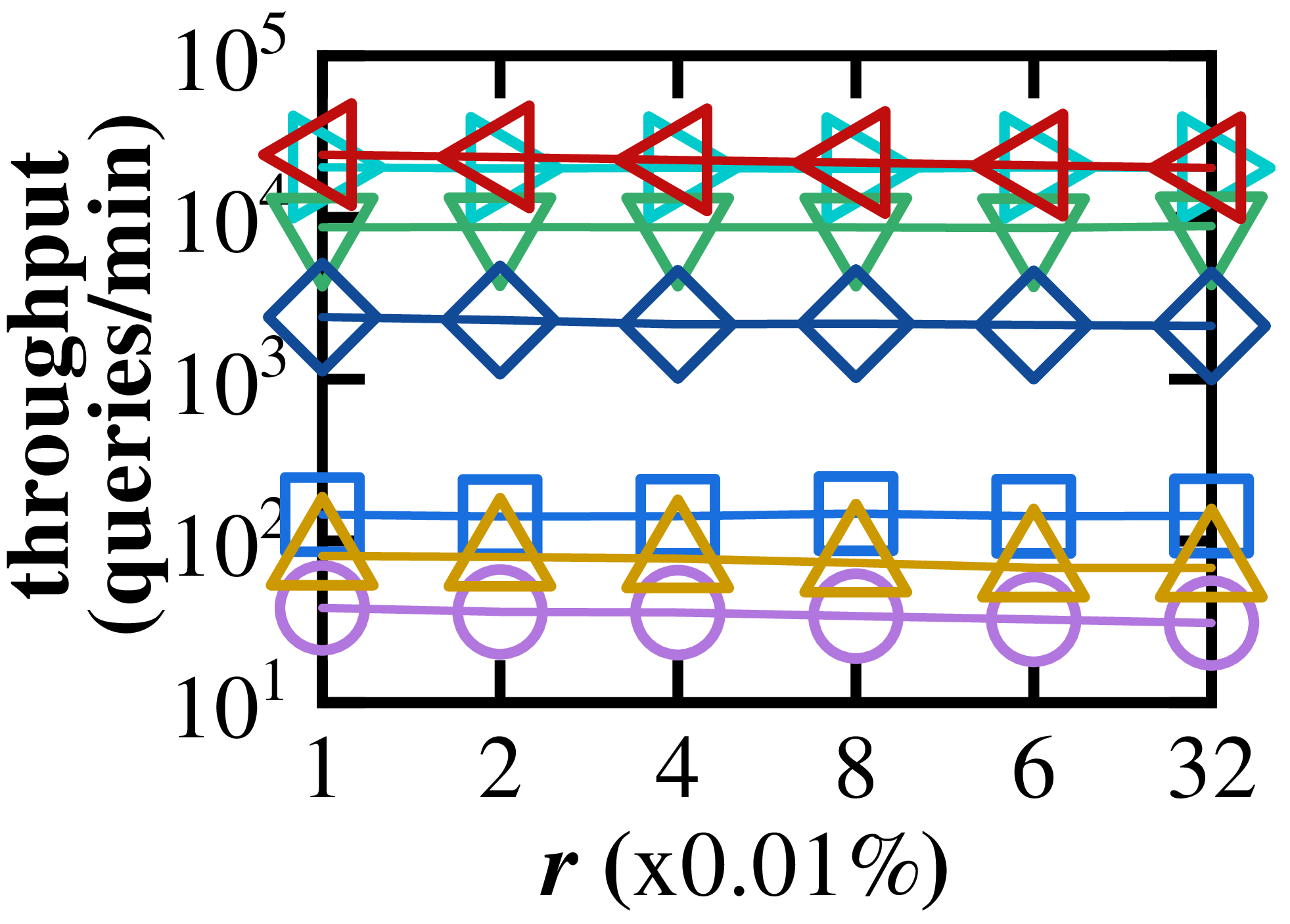}
}
\subfigure[M\textit{k}NNQ on \textit{Words}]{
  \label{fig:rnn-02}
  \includegraphics[width=0.23\textwidth]{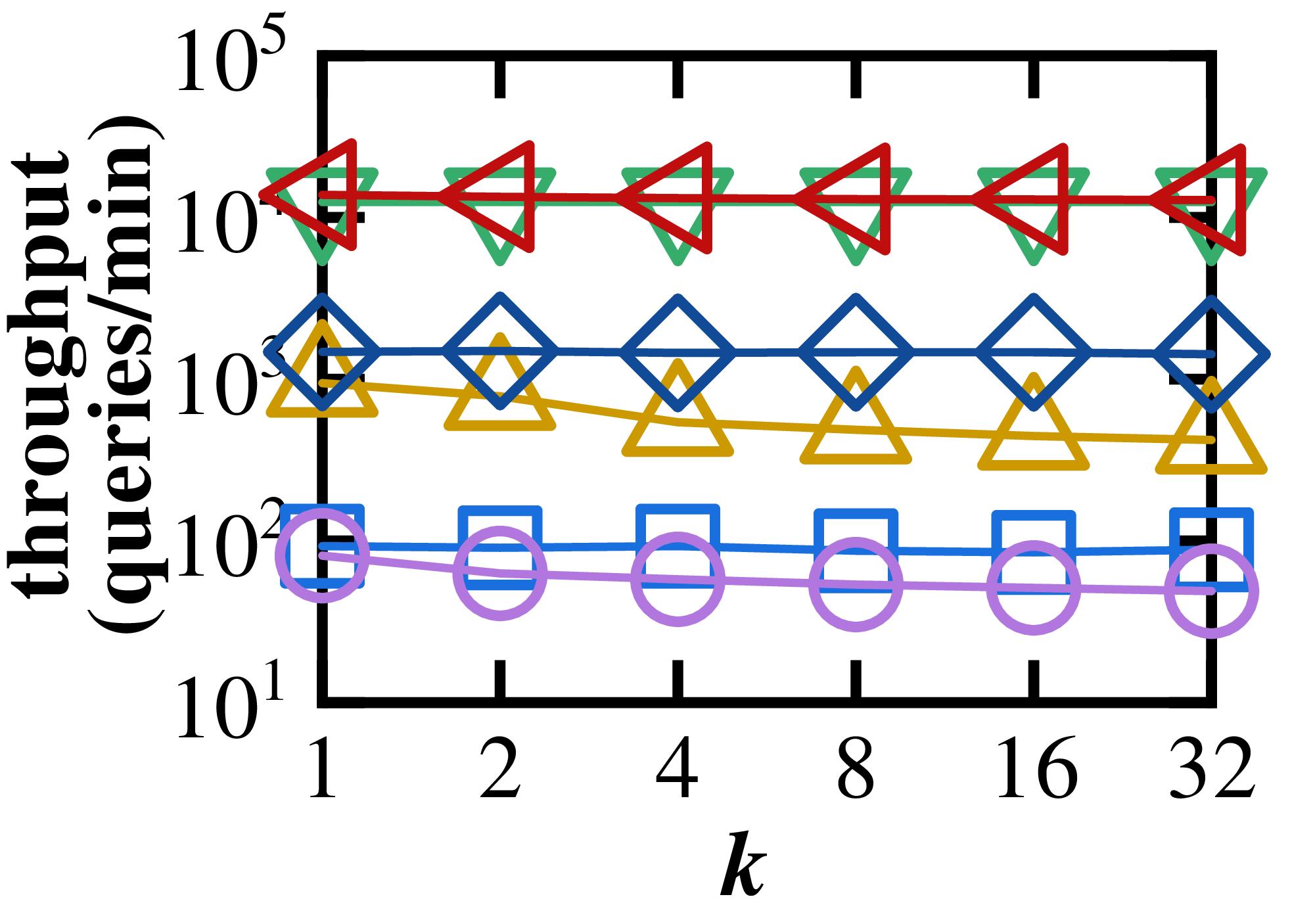}
}
\subfigure[M\textit{k}NNQ on \textit{T-Loc}]{
  \label{fig:rnn-02}
  \includegraphics[width=0.23\textwidth]{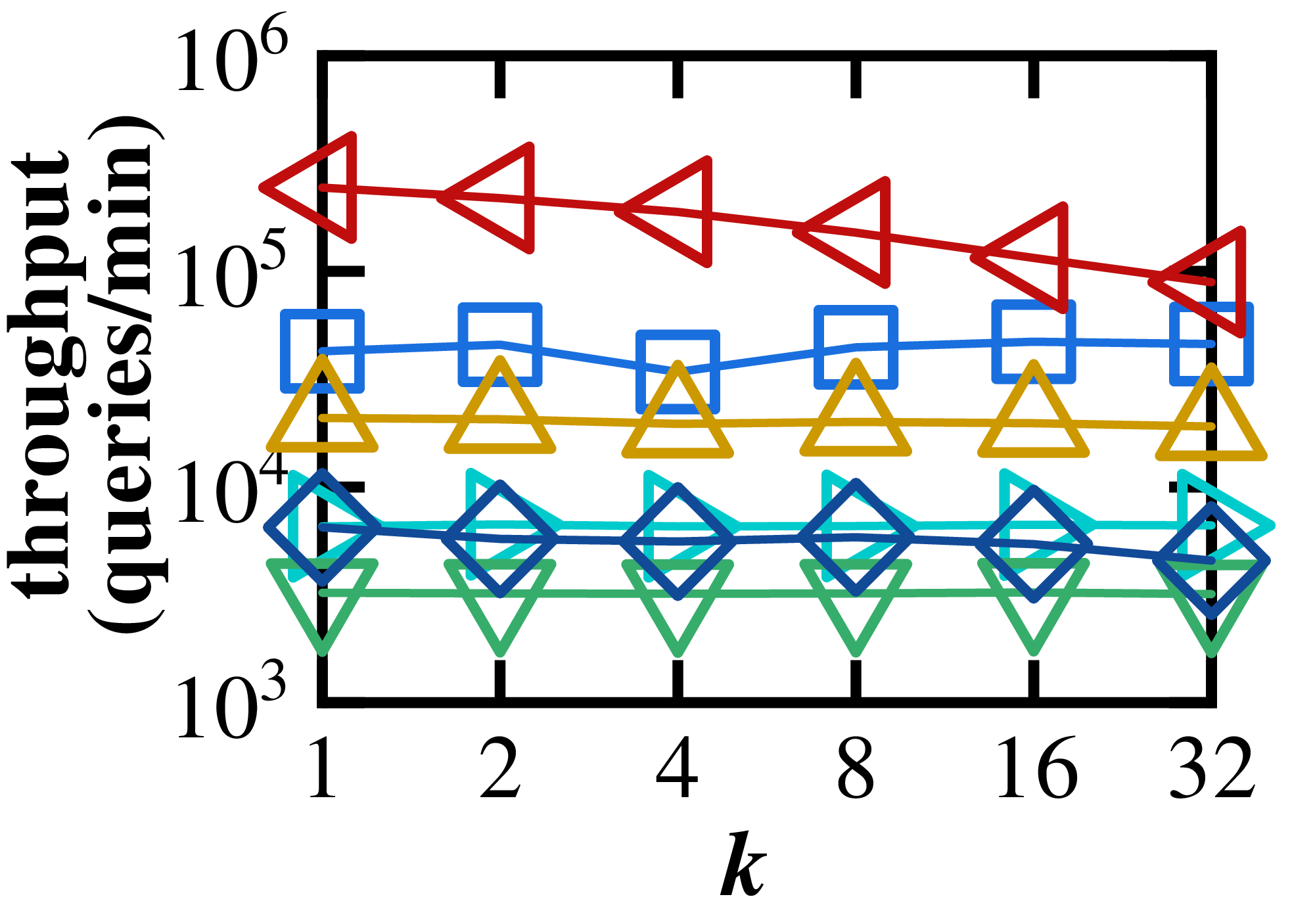}
}
\subfigure[M\textit{k}NNQ on \textit{Vector}]{
  \label{fig:rnn-02}
  \includegraphics[width=0.23\textwidth]{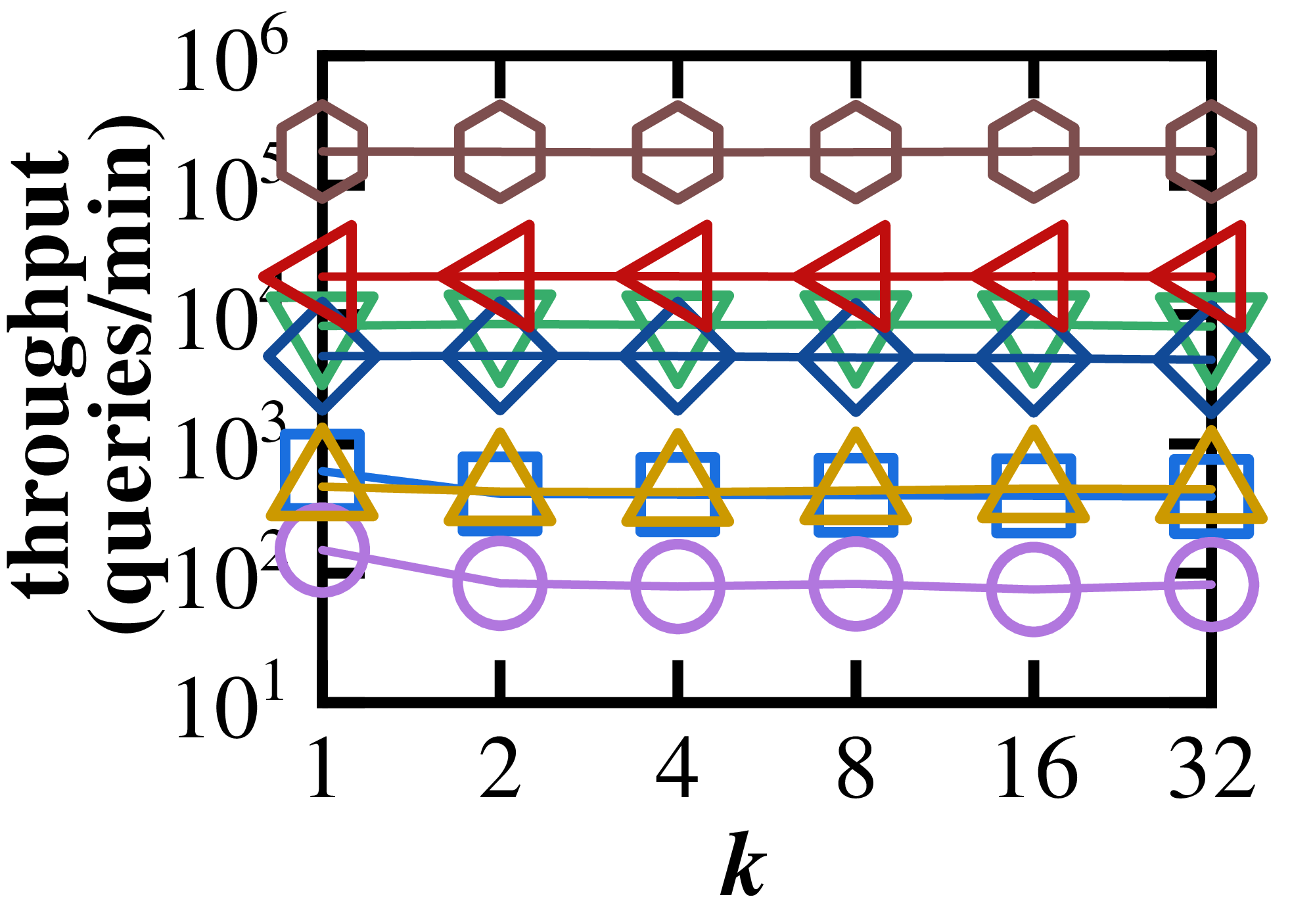}
}

\subfigure[M\textit{k}NNQ on \rev{\textit{DNA}}]{
  \label{fig:rnn-02}
  \includegraphics[width=0.23\textwidth]{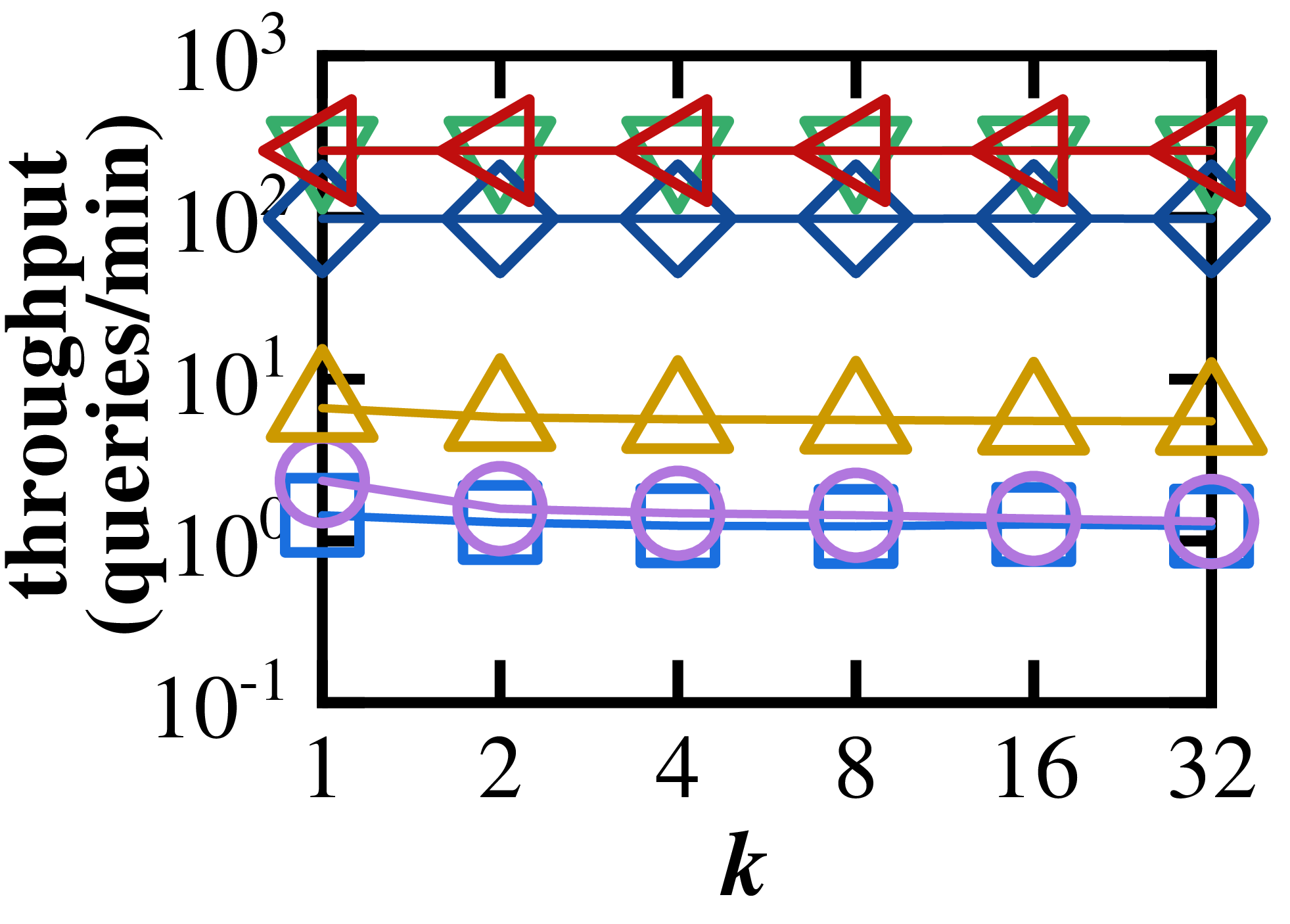}
}
\subfigure[M\textit{k}NNQ on \textit{Color}]{
  \label{fig:rnn-02}
  \includegraphics[width=0.23\textwidth]{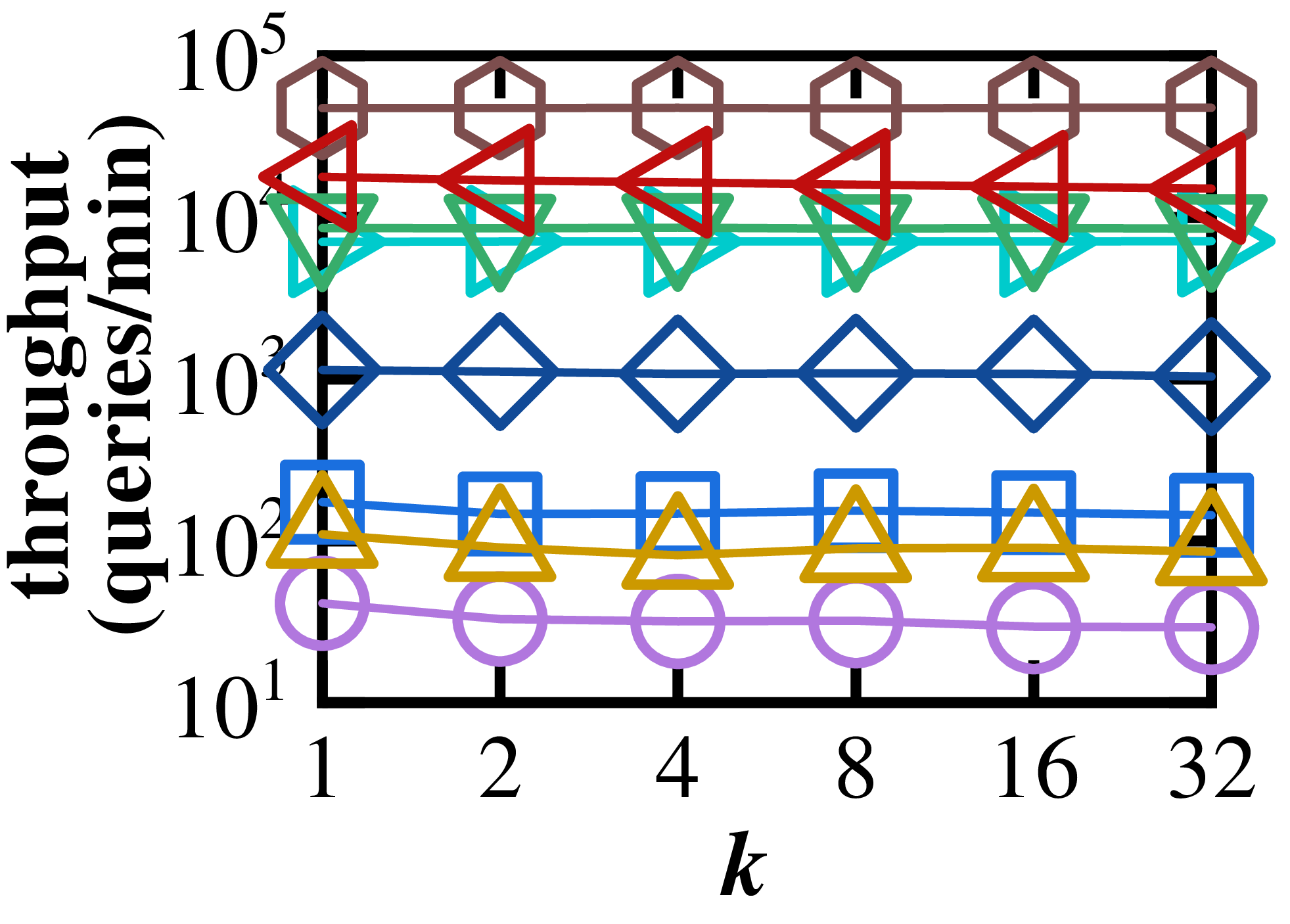}
}
\vspace{-0.4cm}
\caption{\rev{Effect of $r$ and $k$}}
\vspace{-0.5cm}
\label{fig:knnn}
\end{center}
\end{figure*}

\subsection{Construction and Update Performance}
\label{sec:expcons}
Firstly, we conduct a comparative analysis of the construction cost of {\sf GTS} against state-of-the-art competitors. Notably, {\sf GPU-Table} directly constructs one-time distance tables for online queries, eliminating index construction cost. \rev{However, {\sf EGNAT} and {\sf GANNS} encounter memory issues during the index construction for the \textit{T-Loc} dataset, leading to unreported results in this regard.} The results, presented in Table~\ref{tab:construction_cost}, clearly highlight the significantly low construction costs of {\sf GTS}. Notably, \rev{{\sf GTS} completes index construction in less than 3 seconds for each dataset, achieving 1.5$\sim$10$\times$ lower construction time cost compared to other general methods across all datasets.} This advantage is attributed to the efficient utilization of continuous memory allocation for the pivot-based tree index, maximizing the usage of computing resources. \rev{Additionally, in datasets \textit{Color} and \textit{Vector} that exhibit similar distance computational complexity, the space overhead of {\sf GTS} index on \textit{Color} is \rev{almost 4$\times$} that of \textit{Vector}, due to the \rev{fivefold} difference in data volume.  However, the time cost of {\sf GTS} index on \textit{Color} is only \rev{2.7}$\times$ that of \textit{Vector}. This indicates that the parallel computing capability of GPUs on \textit{Vector} exceeds the dataset size, resulting in time complexity overhead lower than $O(n)$. This aligns with the analysis of the proposed construction cost model.} \rev{Furthermore, compared with special-purpose solutions, we observe that: {\bf (i)} {\sf LBPG-Tree} exhibits a lower construction cost, while {\sf GTS} supports more datasets with various distance metrics; and {\bf (ii)} compared to {\sf GANNS}, {\sf GTS} exceeds in constructing a 40 $\times$ smaller index within more than 10$\times$ faster speed, and supports larger datasets, such as \textit{T-Loc}.}

Next, we investigate the impact of the cache table size on update operations. To evaluate the effect of update operations, we perform 5000 updates. Each update operation involves randomly removing an object from {\sf GTS}, subsequently reinserting it, and performing a random similarity range query. Notably, the index is efficiently rebuilt once the number of objects stored in the cache table exceeds its size limit.
The results in Table~\ref{tab:update_cost} demonstrate that {\sf GTS} supports efficient streaming object updates. Notably, the update efficiency of {\sf GTS} exhibits an interesting trend, initially decreasing and then increasing concerning the cache table size, which can be attributed to the trade-off between update efficiency and search efficiency. Specifically, a larger cache table size implies less available space for concurrent similarity search, leading to a higher search cost as fewer queries can be searched simultaneously. To strike a balance between update efficiency and search efficiency, we recommend setting the cache table size to be around 5KB. {In real applications, the cache size could be flexibly adjusted based on our proposed index updating cost model to meet users’ requirements.} 

We also compare the time cost of streaming data updates and batch data updates. For the former, we simulate the process of randomly removing an object and then reinserting it. For the latter, we remove 10$\%$ of the objects from the dataset randomly and then insert them back. \rev{The results, presented in Fig.~\ref{fig:exp-update}, reveal that CPU-based methods exhibit higher efficiency in handling streaming data updates, while GPU-based methods demonstrate higher efficiency in managing batch updates for large and complex data. Notably, {\sf GTS} showcases remarkable performance for streaming data updates compared to other GPU-based methods, including {\sf LBPG-Tree} and {\sf GANNS}. These alternatives necessitate a complete rebuild from scratch for any data updates. Additionally, {\sf GPU-Tree}, leveraging single GPU cores for complex tree structure updating, faces an efficiency bottleneck. Moreover, {\sf GTS} emerges as the optimal solution for managing dynamic data updates on {\textit{DNA}}. These findings imply that {\sf GTS} is particularly well-suited for handling diverse cancer omics data with a large volume.}

\subsection{Similarity Search Performance}
\label{sec:ssp}
We proceed to explore the similarity search performances of {\sf GTS} and its competitors by varying three parameters: the tuning parameter node capacity, the search radius $r$ for MRQ, and the desired number $k$ for M\textit{k}NNQ.

\noindent
\textbf{Effect of Node Capacity $N_c$.} Fig.~\ref{fig:exp-partition} shows the {\sf GTS}'s performance of MRQ and M\textit{k}NNQ under various node capacities $N_c$. \rev{Generally, the query efficiency exhibits alternatively increasing and decreasing patterns. A small $N_c$ requires visiting more tree layers, reducing the level of computing concurrency. On the other hand, a small $N_c$ also utilizes more pivots for object pruning, thereby avoiding a higher volume of unnecessary distance computations. Consequently, search performance depends on the trade-off between parallelism and pruning efficiency, while our evaluation validates that a small $N_c$ can strike a balance and showcase high search performance, which is consistent with our analysis based on the proposed cost model presented in Section~\ref{sec:cost}}. Therefore, we set the node capacity to 20 for higher search efficiency.

\begin{figure}[t]
\begin{center}
\subfigtopskip=-1pt
\subfigcapskip=-1pt
\subfigure[\textit{T-Loc}]{  
  \includegraphics[width=0.3\textwidth]{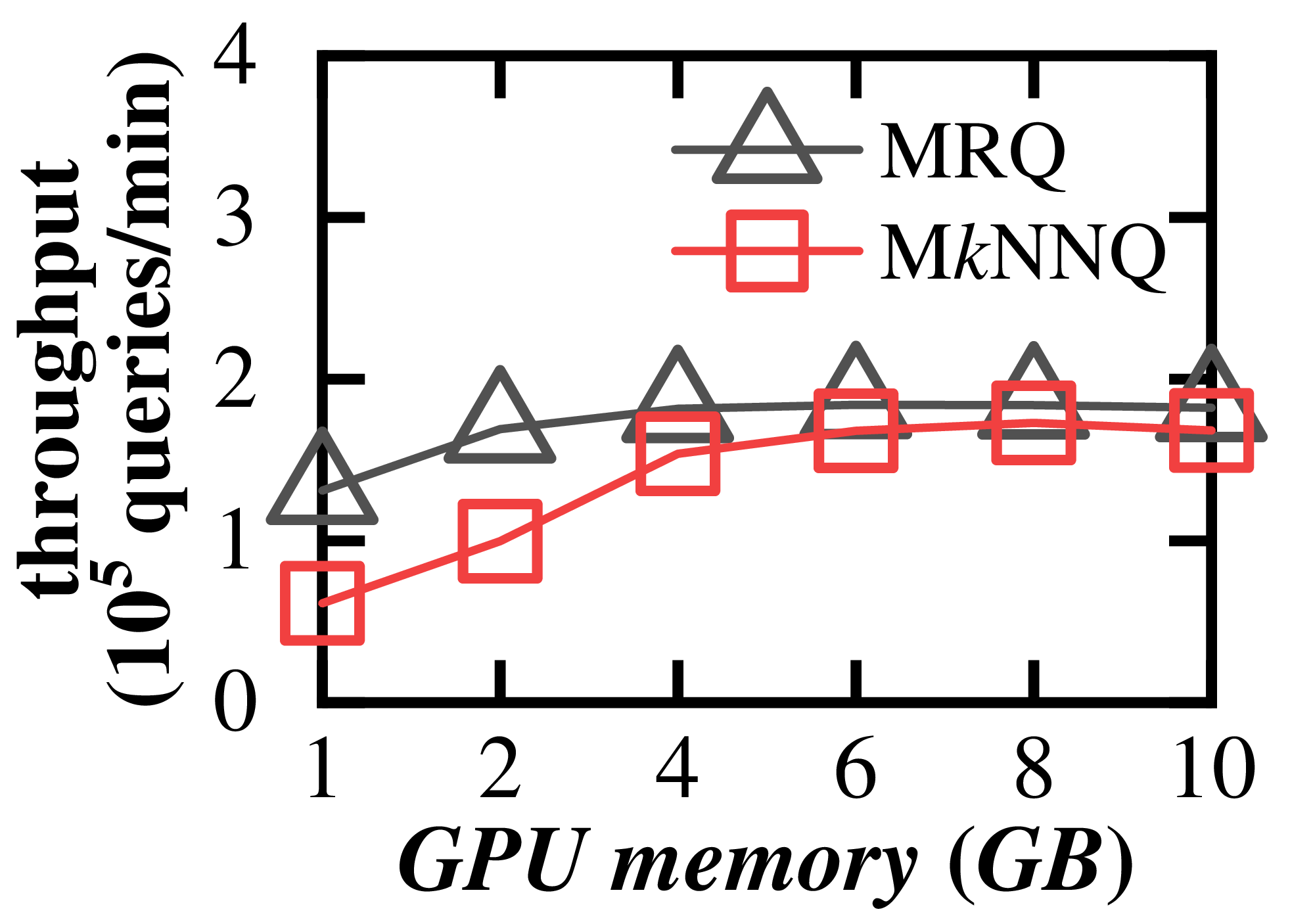}  
}
\subfigure[\textit{Color}]{  
  \includegraphics[width=0.3\textwidth]{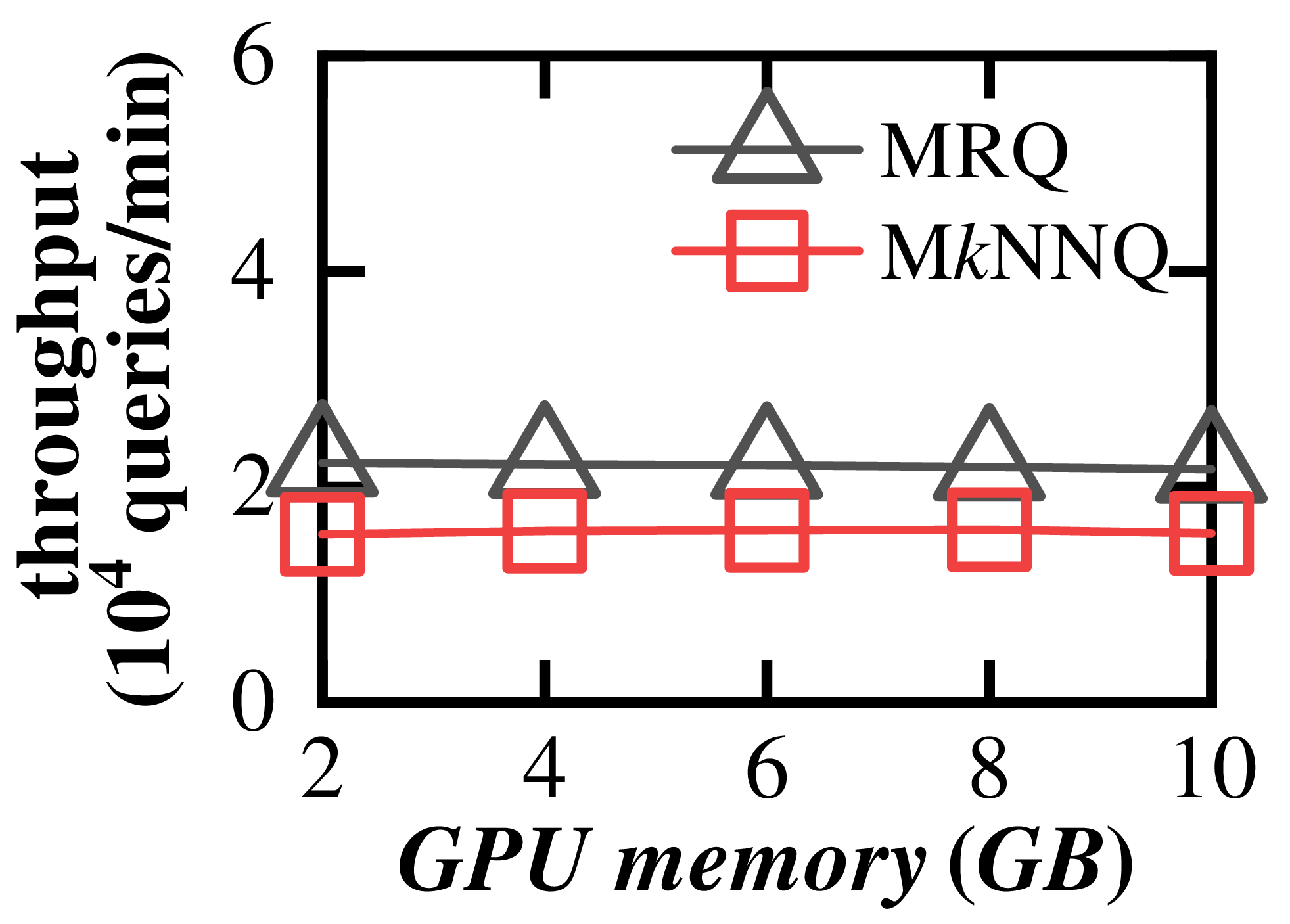}  
}
\vspace{-0.4cm}
\caption{\rev{Effect of the GPU memory}}
\vspace{-0.4cm}
\label{fig:exp-memory}
\end{center}
\end{figure}


\noindent
\textbf{Effects of $r$ and $k$.} In Fig.~\ref{fig:knnn}, we present the performance of MRQ and  M\textit{k}NNQ for different algorithms on \rev{five} real datasets, varying the search radius $r$ and the integer $k$. Our proposed GPU-based tree index {\sf GTS} demonstrates superior efficiency, outperforming \rev{all baseline methods for general metric space} across all datasets. This demonstrates the effectiveness and efficiency of our concurrent similarity search algorithms. Notably, {\sf GTS} achieves up to two orders of magnitude higher throughput performance compared to CPU-based methods and supports up to 20x larger throughput than \rev{GPU-based general methods}. This significant speedup can be attributed to two main factors. First, the proposed tree-based index stored in a continuous table reduces unnecessary distance computations while harnessing the superior computing power of the GPU, resulting in a faster query process. Second, {\sf GTS} can simultaneously answer multiple queries using a two-stage concurrent similarity search, effectively managing the memory cost for concurrent queries to optimize GPU computing resources. As a result, {\sf GTS} dramatically improves the similarity search performance. 

\rev{Compared with special-purpose solutions, we find that: {\bf (i)} {\sf GTS} achieves higher efficiency on all the datasets than {\sf LBPG-Tree}, due to the better usage of GPU computing power; and {\bf (ii)} though {\sf GANNS} outperforms {\sf GTS} in M\textit{k}NNQ efficiency measured in the order of milliseconds, {\sf GTS} supports a broader range of datasets and offers efficient MRQ capabilities. This suggests that {\sf GTS} has wider applicability and can be utilized in various scenarios.}

\begin{figure}[t]
\begin{center}
\subfigtopskip=-7pt
\subfigcapskip=-3pt
\includegraphics[height=0.5cm]{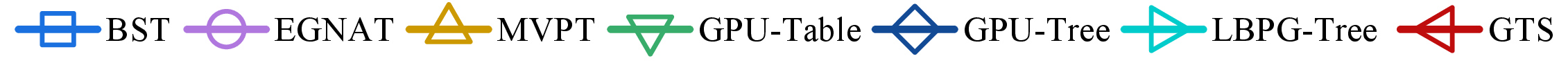}\vspace{0.2cm}

\subfigure[\textit{T-Loc}]{
  \label{fig:rnn-02}
  \includegraphics[width=0.3\textwidth]{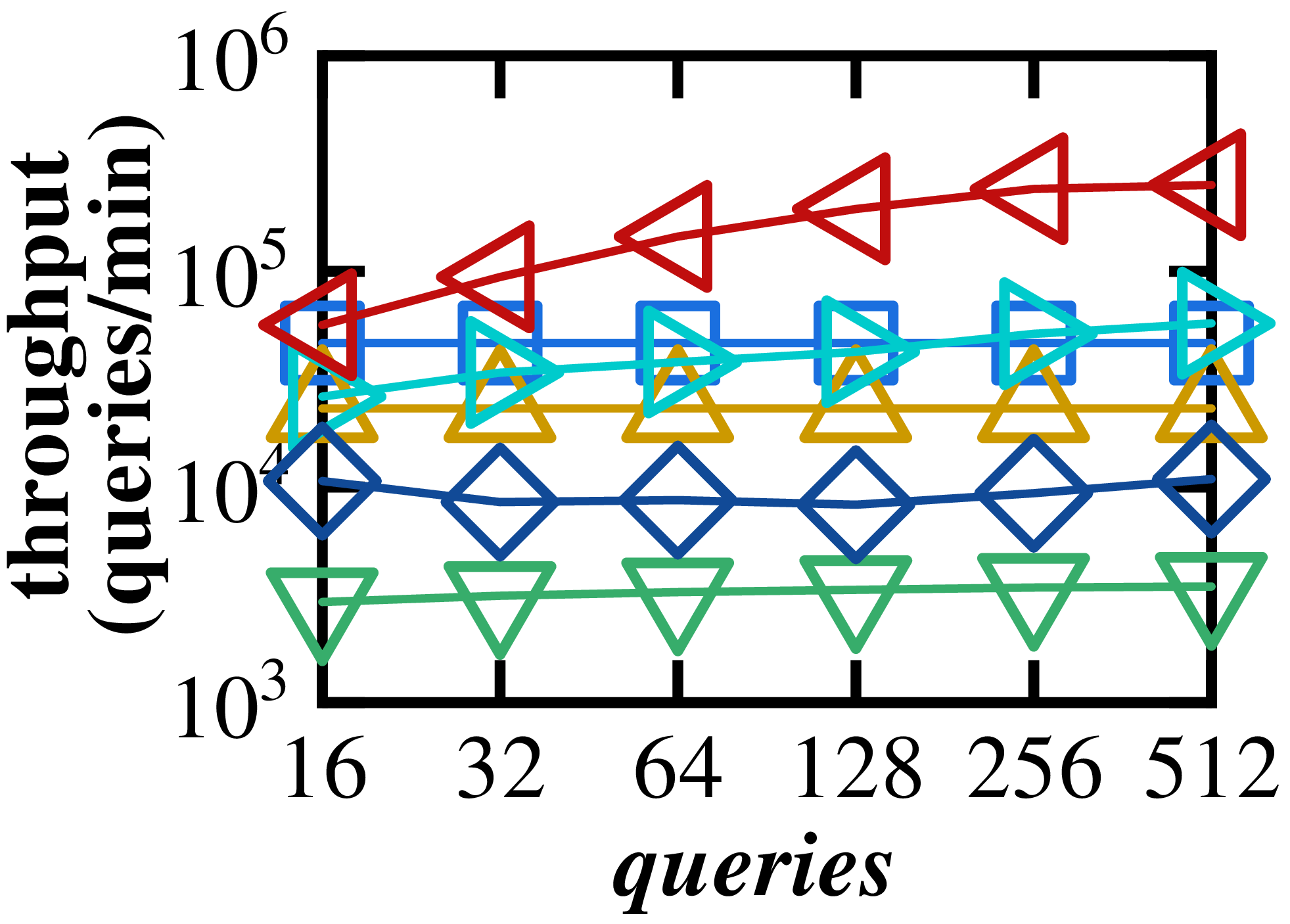}
}
\subfigure[\textit{Color}]{
  \label{fig:rnn-02}
  \includegraphics[width=0.3\textwidth]{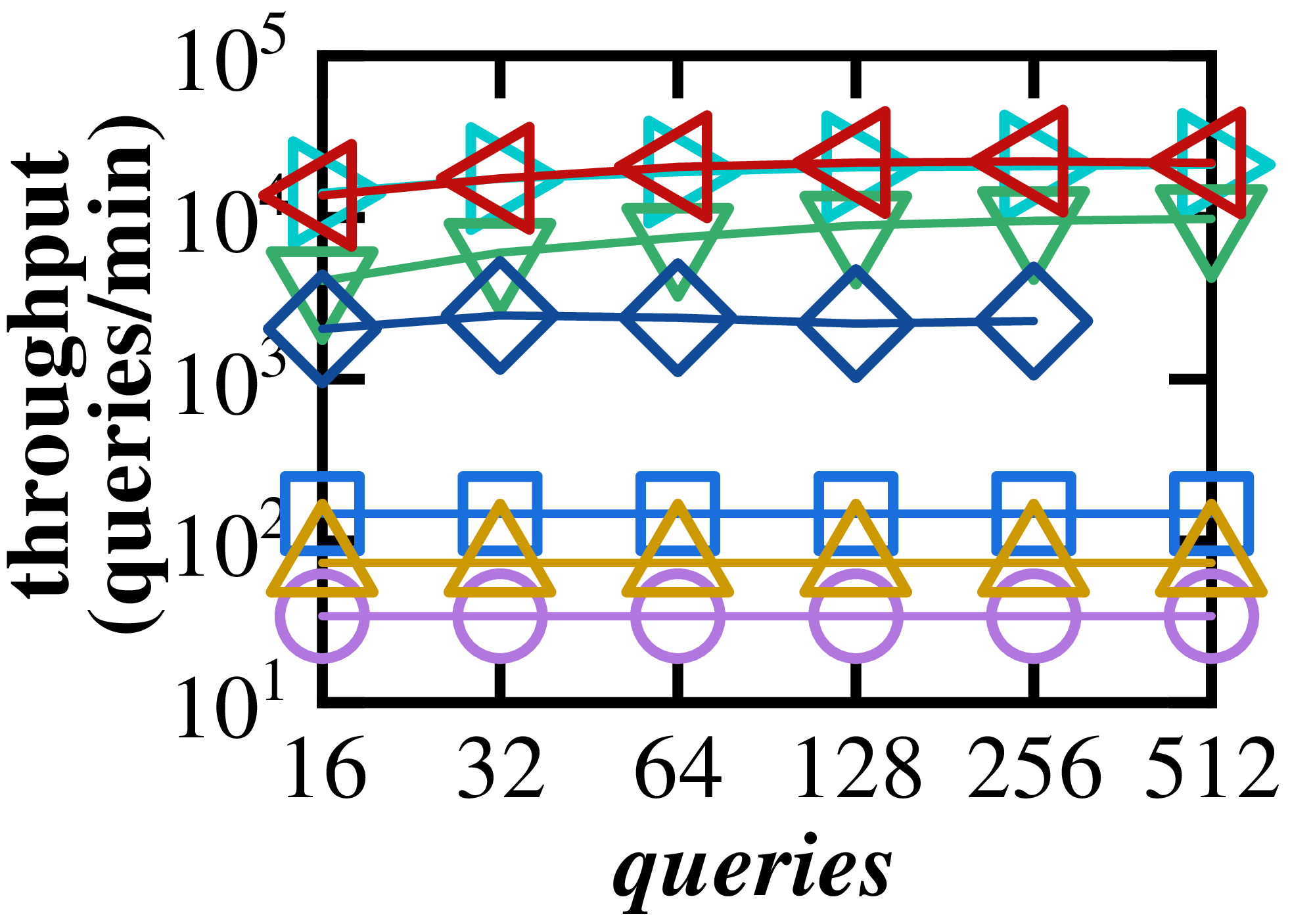}    
}
\vspace{-0.4cm}
\caption{\rev{MRQ performance vs. the number of queries}}
\vspace{-0.4cm}
\label{fig:query-knn}
\end{center}
\end{figure}

\subsection{Scalability Analysis}
Finally, we access the scalability of {\sf GTS} by varying {the GPU memory,} the number of queries in a \rev{batch}, \rev{the percentage of distinct objects}, and dataset cardinality.

\noindent
\textbf{{Effect of {GPU memory}.}} 
{In Fig.~\ref{fig:exp-memory}, we present the results for M\textit{k}NNQ and MRQ when varying the GPU memory size. \rev{Due to the dataset size of \textit{Color} is 1.1G, the results of \textit{Color} using 1G GPU memory is not reported.} As observed, \rev{the throughput} generally increases when more GPU memory becomes available. Notably, the throughput remains unchanged with the increasing GPU memory on \textit{Color}. \rev{The reason behind this is that GPU computing resources are fully utilized, while only a small part of memory is sufficient for all the computing tasks, making the impact of GPU memory limited under such conditions.}


\noindent
\textbf{{Effect of the {Concurrency}.}} 
To demonstrate the effectiveness of our proposed similarity batch search strategy, we compare the M\textit{k}NNQ performance of {\sf GTS} with its competitors while varying the number of queries in a \rev{batch} in Fig.~\ref{fig:query-knn}. \rev{Firstly, we observe that {\sf GPU-Tree} faces the memory deadlock problem on \textit{Color} for 512 queries in a batch, justifying our analysis in Section 1.} Besides, with the increasing number of queries, the \rev{throughput} of all GPU-based methods rises as more similarity measurements can be computed in parallel. 
In contrast, CPU-based methods remain unaffected by the number of queries. Remarkably, {\sf GTS} consistently outperforms other methods, demonstrating its superior ability to support concurrent similarity search.

\begin{figure}[t]
\begin{center}
\subfigtopskip=-2pt
\subfigcapskip=-1pt
\subfigure[\textit{T-Loc}]{
  \label{fig:rnn-02}
  \includegraphics[width=0.3\textwidth]{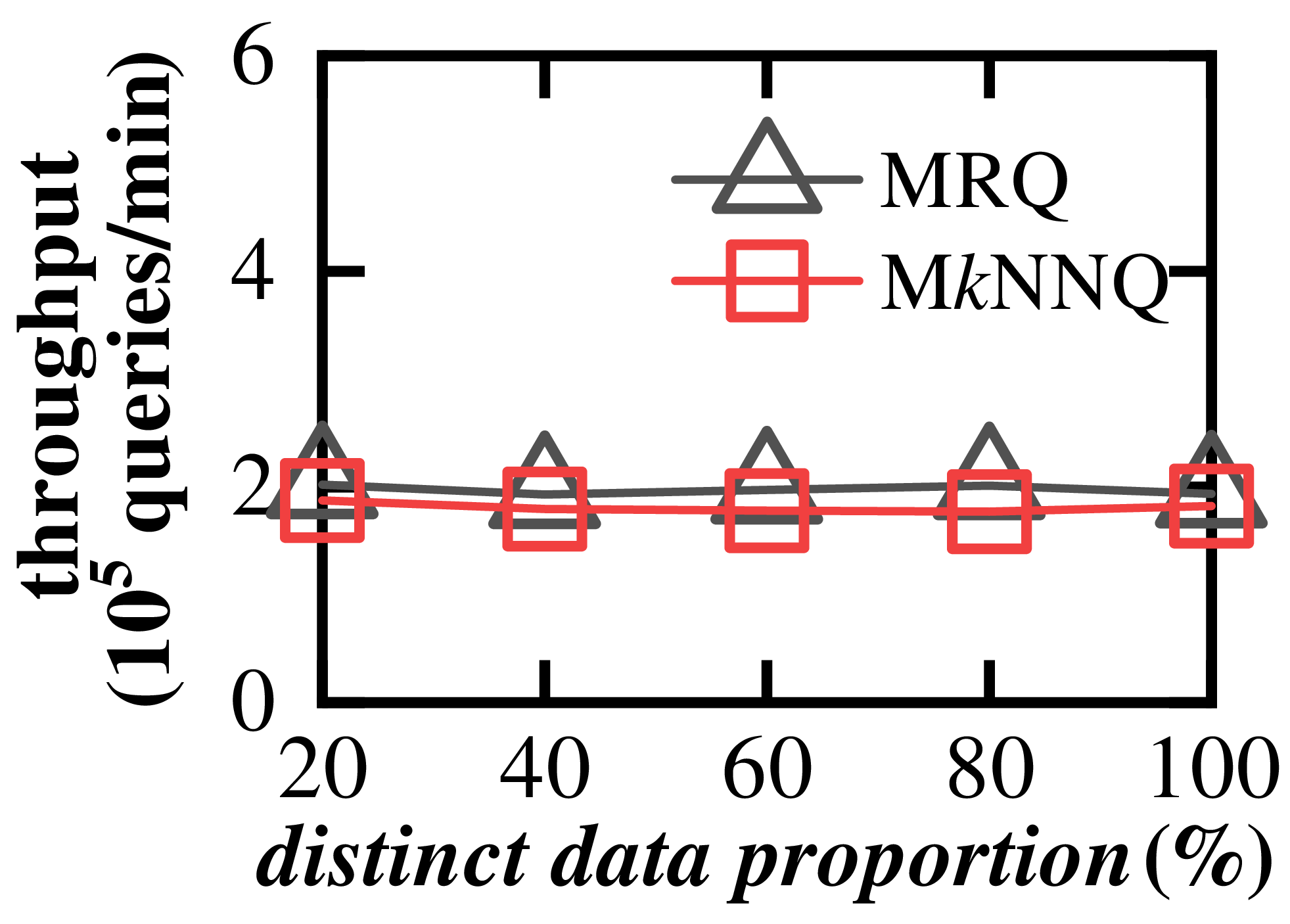}  
}
\subfigure[\textit{Color}]{
  \label{fig:rnn-02}
  \includegraphics[width=0.3\textwidth]{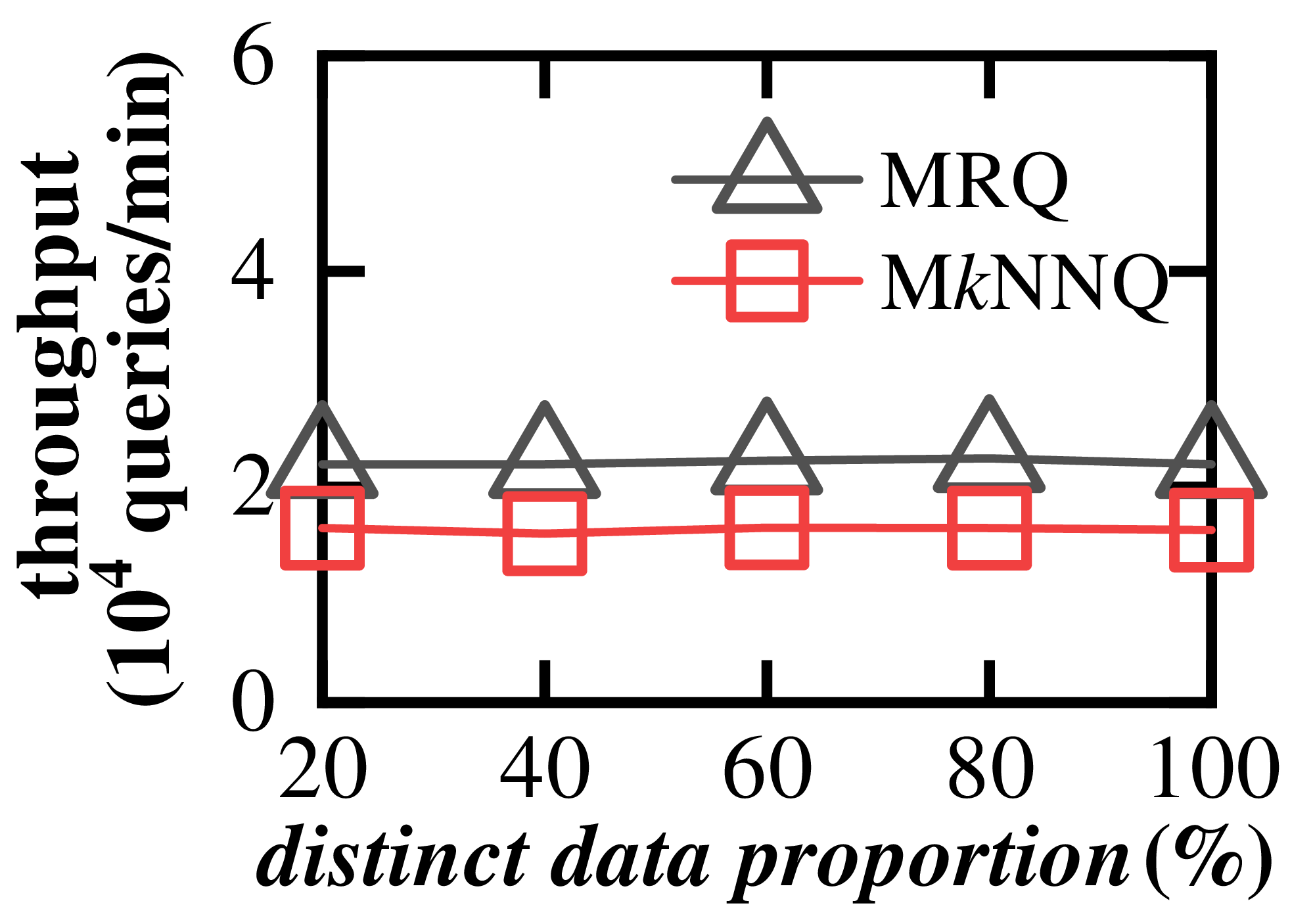}  
}
\vspace{-0.4cm}
\caption{\rev{Effect of identical objects}}
\vspace{-0.4cm}
\label{fig:exp-distinct}
\end{center}
\end{figure}

\vspace{0.2cm}
\noindent
\textbf{\rev{Effect of {identical objects}.}} 
\rev{To investigate the influence of identical objects on the efficiency of similarity search, we conduct experiments with varying proportions of distinct data on the \textit{T-Loc} and \textit{Color} datasets. The findings presented in Fig.~\ref{fig:exp-distinct} illustrate that the search performance of the proposed {\sf GTS} remains unaffected by the presence of identical objects, thereby validating the effectiveness of our approach.}

\vspace{0.2cm}
\noindent
\textbf{Effect of Dataset Cardinality.} We vary the cardinality of all datasets from 20\% to 100\% and present the M\textit{k}NNQ results in Fig.~\ref{fig:card-rnn}.   
\rev{As observed, the throughput decreases with the dataset size. Notably, the increasing dataset size poses a memory issue for {\sf EGNAt} on \textit{T-Loc} and \textit{Color}, due to the large storage size for storing pre-computed distances. Meanwhile, some GPU-based methods, including {\sf GPU-Tree} and {\sf GANNS}, also face memory issue on \textit{Color}. 
This is attributed to their utilization of GPU-cores for various queries, demanding a significant amount of memory to store intermediate results. Meanwhile, due to {\sf LBPG-Tree} facing the dimension curse, it also runs out of memory on \textit{Color} with a cardinality of 80\%. Conversely, {\sf GTS} scales well with the increasing dataset size. This is because of our proposal to dynamically group the intermediate results of each query according to the available memory. We then process each group sequentially to obtain real results, with the queries in each group computed in parallel, circumventing memory deadlocks.} Thus, {\sf GTS} scales effectively with increasing data sizes.

\vspace{0.2cm}
\noindent
\textbf{Remark.} \rev{Throughout the entire experiment, {\sf GTS} consistently outperforms general methods, and stands out as the optimal solution for \textit{Words}, \textit{T-loc}, and \textit{DNA}. Based on these results, we can conclude that our GPU-based tree index {\sf GTS} holds the promise for efficiently managing dynamic data of various types with flexible metrics. It proves to be a pivotal solution for enhancing the scalability of vector databases and offering cost-effective updates.}

\begin{figure}[t]
\begin{center}
\subfigtopskip=-2pt
\subfigcapskip=-1pt
\includegraphics[height=0.4cm]{ExpFigs/icon-l.eps}\vspace{0.2cm}

\subfigure[\textit{T-Loc}]{
  \label{fig:rnn-02}
  \includegraphics[width=0.23\textwidth]{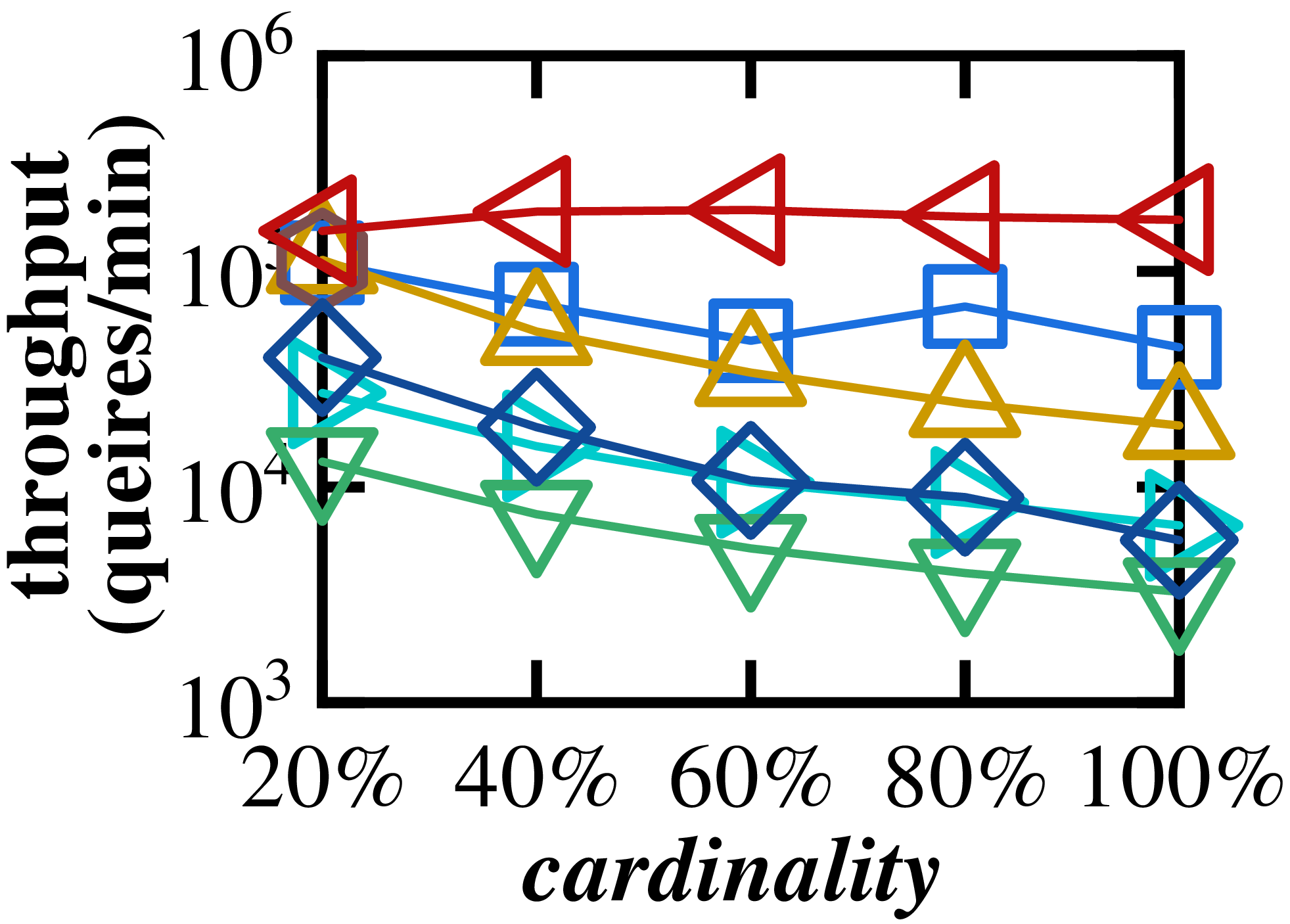}  
}
\subfigure[\textit{T-Loc}]{
  \label{fig:rnn-02}
  \includegraphics[width=0.23\textwidth]{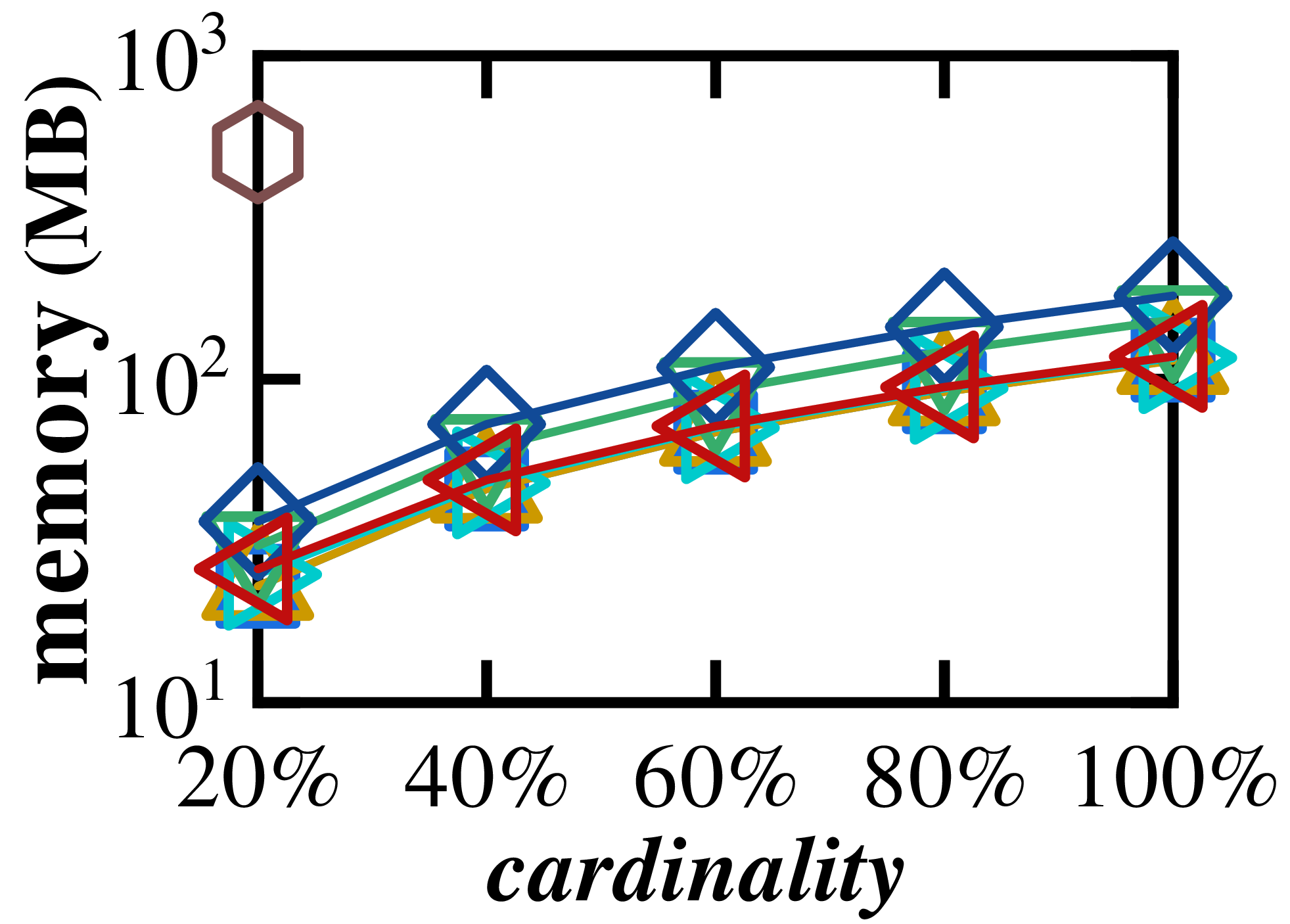}  
}
\subfigure[\textit{Color}]{
  \label{fig:rnn-02}
  \includegraphics[width=0.23\textwidth]{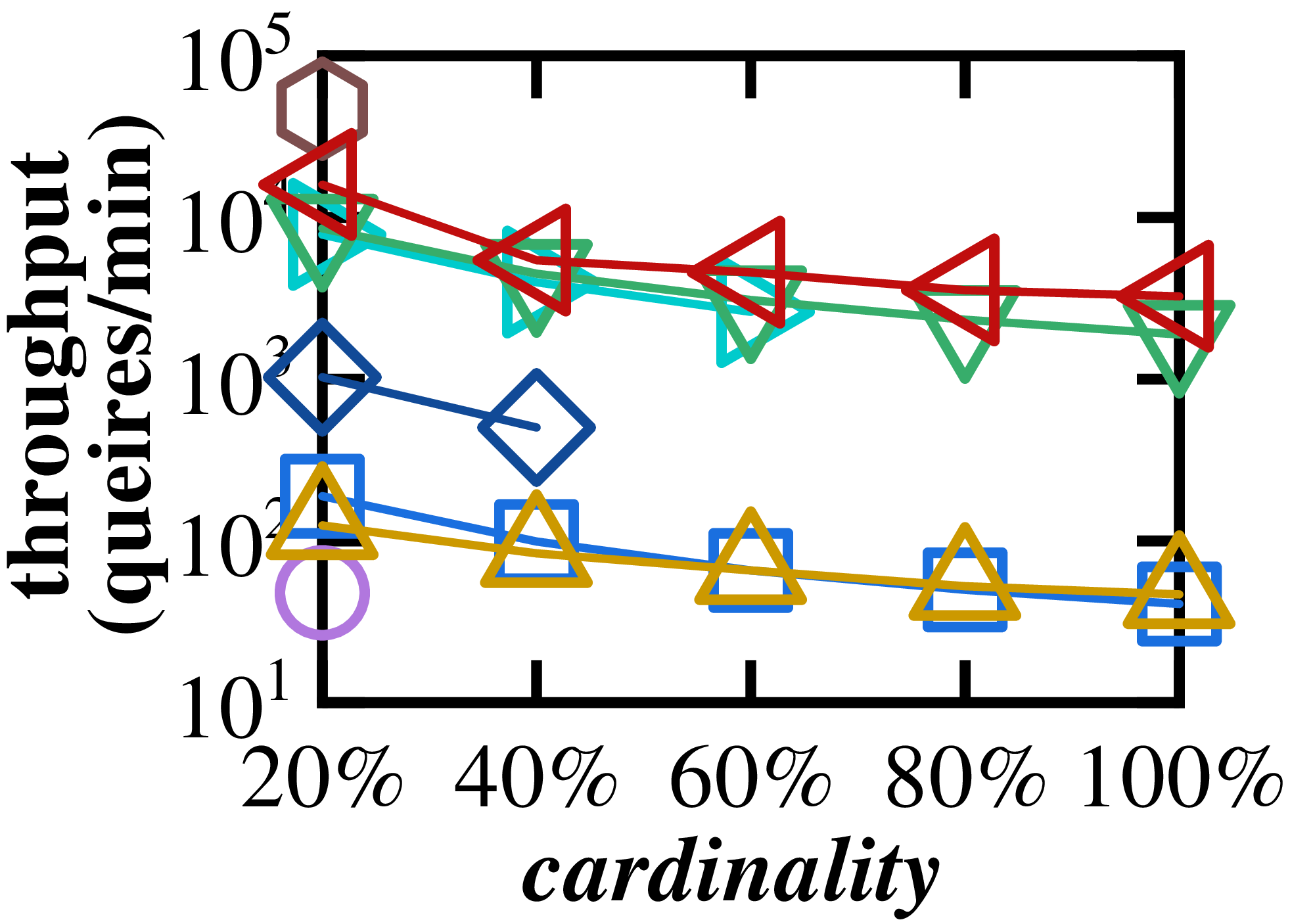}  
}
\subfigure[\textit{Color}]{
  \label{fig:rnn-02}
  \includegraphics[width=0.23\textwidth]{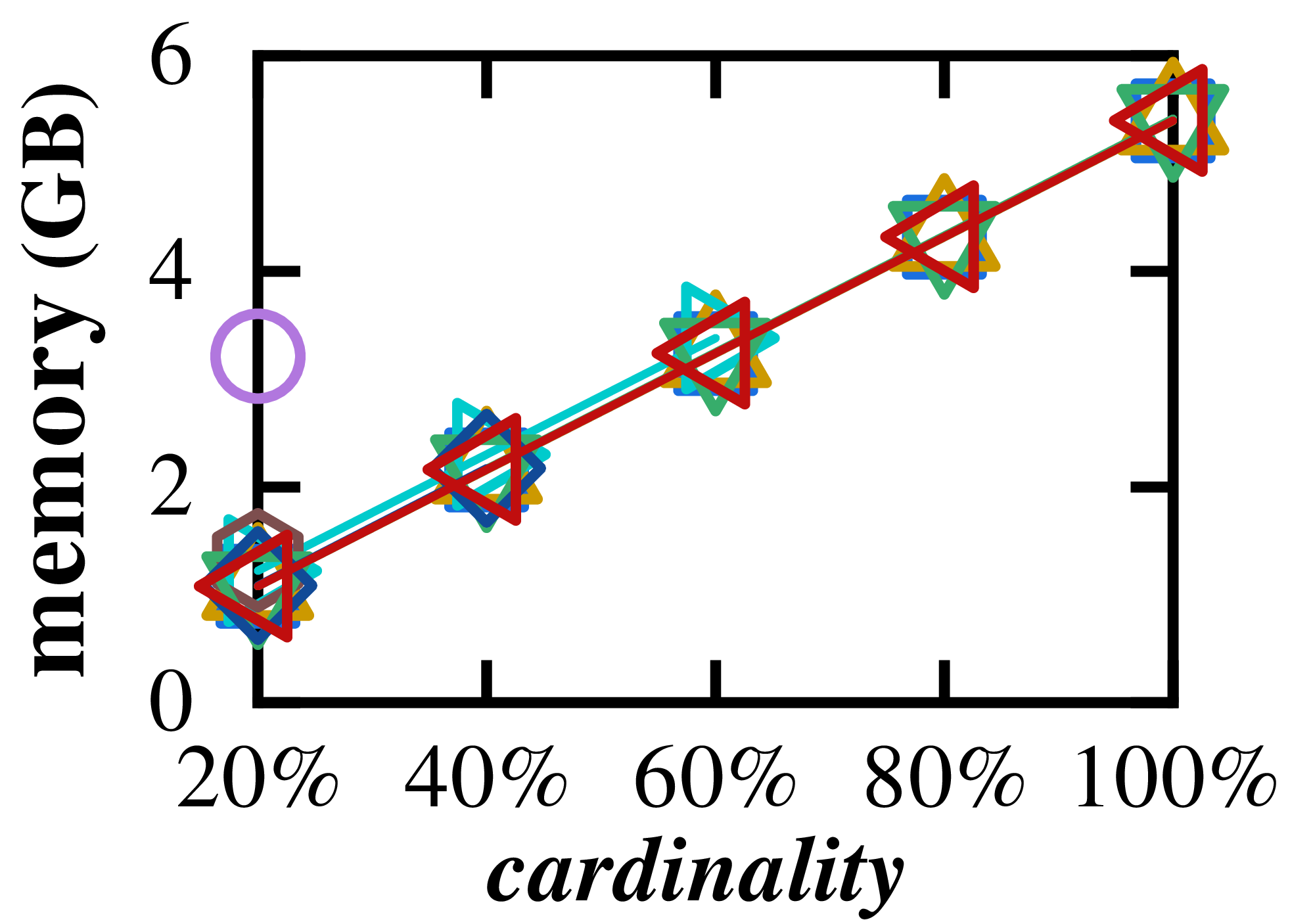}  
}

\vspace{-0.4cm}
\caption{\rev{M\textit{k}NNQ performance consumption vs. \textit{Cardinality}}}
\vspace{-0.5cm}
\label{fig:card-rnn}
\end{center}
\end{figure}

\section{conclusions}
\label{sec:conclusion}
In this paper, we propose {\sf GTS}, a highly efficient GPU-based tree index for similarity search. {\sf GTS} incorporates a  pivot-based tree index for object management, stored in a continuous node list, along with a cache table to support streaming data updates and batch updates for dynamic scenarios. In addition, we introduce efficient two-stage search methods that support concurrent similarity search with high query throughput. To strike a balance between the pruning capability and the parallel computing efficiency, we develop a cost-based optimization model. Extensive experiments demonstrate that our {\sf GTS} outperforms state-of-the-art CPU-based and GPU-based methods, offering more efficient concurrent similarity search and scaling well with data size.
These results highlight the superior efficiency and scalability of {\sf GTS}, making it a promising solution for various real-life applications. Moving forward, we plan to explore approximate similarity search with learned index based on GPU architecture to further enhance efficiency.

\begin{acks}
This work was supported in part by the NSFC under Grants No. (62025206, U23A20296, and 62102351),
Zhejiang Province's "Lingyan" R\&D Project under Grant No. 2024C01259, Yongjiang Talent Introduction Programme (2022A-237-G), and the Key
Lab of Intelligent Computing Based Big Data of Zhejiang Province. Yunjun Gao is the corresponding author of the work.

\end{acks}

\bibliographystyle{ACM-Reference-Format}
\bibliography{reference}

\end{document}